\documentclass[twocolumn]{aastex63}
\usepackage{natbib}

\newcommand{\normaldiff}[2]{\displaystyle \frac{d #1}{d #2}}
\newcommand{\partialdiff}[2]{\displaystyle \frac{\partial #1}{\partial #2}}
\newcommand\Msun{M_\odot}

\newcommand\Msunperyear{\ M_\odot \ {\rm yr}^{-1}}
\newcommand\pow[1]{$^{#1}$}
\defcitealias{2013ApJ...778L..23T}{TLM13}

\received{December, 6, 2021}
\accepted{April, 14, 2022}
\submitjournal{ApJ}

\shorttitle{Evolution of a USSNR}
\shortauthors{Matsuoka et al.}

\begin{document}

\title{Long-term evolution of a supernova remnant hosting a double neutron star binary}

\correspondingauthor{Tomoki Matsuoka}
\email{t.matsuoka@kusastro.kyoto-u.ac.jp}

\author[0000-0002-6916-3559]{Tomoki Matsuoka}
\affil{Department of Astronomy, Kyoto University, Kitashirakawa-Oiwake-cho, Sakyo-ku, Kyoto, 606-8502, Japan}

\author[0000-0002-2899-4241]{Shiu-Hang Lee}
\affiliation{Department of Astronomy, Kyoto University, Kitashirakawa-Oiwake-cho, Sakyo-ku, Kyoto, 606-8502, Japan}
\affiliation{Kavli Institute for the Physics and Mathematics of the Universe (WPI), The University of Tokyo, Kashiwa 277-8583, Japan}

\author[0000-0003-2611-7269]{Keiichi Maeda}
\affiliation{Department of Astronomy, Kyoto University, Kitashirakawa-Oiwake-cho, Sakyo-ku, Kyoto, 606-8502, Japan}

\author[0000-0003-0304-9283]{Tomoya Takiwaki}
\affiliation{National Astronomical Observatory of Japan, 2-21-1, Osawa, Mitaka, Tokyo, 181-8588, Japan}

\author[0000-0003-1169-1954]{Takashi J. Moriya}
\affiliation{National Astronomical Observatory of Japan, 2-21-1, Osawa, Mitaka, Tokyo, 181-8588, Japan}
\affiliation{School of Physics and Astronomy, Faculty of Science, Monash University, Clayton, Victoria 2800, Australia}

\begin{abstract}
An ultra-stripped supernova (USSN) is a type of core-collapse SN explosion proposed to be a candidate formation site of a double neutron star (DNS) binary. We investigate the dynamical evolution of an ultra-stripped supernova remnant (USSNR), which should host a DNS at its center. By accounting for the mass-loss history of the progenitor binary using a model developed by a previous study, we construct the large-scale structure of the {circumstellar medium (CSM)} up to a radius $\sim 100\,{\rm pc}$, and simulate the explosion and subsequent evolution of a USSN surrounded by such a CSM environment. We find that the CSM encompasses an extended region characterized by a hot plasma with a temperature $\sim 10^8\,$K located around the termination shock of the wind from the progenitor binary ($\sim 10\,$pc), and the USSNR blastwave is drastically weakened while penetrating through this hot plasma. Radio continuum emission from a young USSNR is sufficiently bright to be detectable if it inhabits our Galaxy but faint compared to the observed Galactic SNRs, and thereafter declines in luminosity through adiabatic cooling. Within our parameter space, USSNRs typically exhibit a low radio luminosity and surface brightness compared to the known Galactic SNRs. Due to the small event rate of USSNe and their relatively short observable lifespan, we calculate that USSNRs account for only $\sim 0.1$--$1$ \% of the total SNR population. This is consistent with the fact that no SNR hosting a DNS binary has been discovered in the Milky Way so far.
\end{abstract}

\keywords{supernova remnant}

\section{Introduction}\label{sec:introduction}

A {double neutron star (DNS)} binary is believed to be the fossil object from a binary system of two massive stars which have both exploded as core-collapse supernovae (SNe) in the past \citep[e.g.,][]{2005MNRAS.361.1243P}. Observations of Galactic radio pulsars have revealed that some DNS binaries are in an orbit tight enough to merge within the cosmic age \citep{2003Natur.426..531B}. {Indeed, previous observations for the short gamma-ray burst GRB 130603B have implied the association between the gamma-ray emission and kilonova in the DNS merger \citep{2013Natur.500..547T, 2013ApJ...775..113T, 2013ApJ...778L..16H}.} Furthermore, recent gravitational wave detectors and rapid follow-up electromagnetic observations have succeeded in probing the coalescence of a DNS, confirming the link of these objects to the origin of short gamma-ray bursts and the nucleosynthesis of r-process elements \citep[e.g.,][]{TheLIGOScientific:2017qsa, Monitor:2017mdv,2017PASJ...69..102T}. 

The formation of a DNS requires that the binary system is not disrupted by the evolution history of the massive stars all the way through their core-collapses.
One of the plausible scenarios of DNS formation invokes an ultra-stripped supernova \citep[USSN,][]{2017ApJ...846..170T, 2017MNRAS.471.4275Y}. In a close binary consisting of two massive stars, the primary star first explodes as a SN. After a phase as a high-mass X-ray binary, the outer layer of the secondary star is stripped away in two steps: (1) the ejection of its hydrogen-rich envelope through a phase of common envelope (CE) interaction, and (2) the stripping of the helium layer through Roche lobe overflow (RLO). These binary interactions lead to the formation of an helium star $(\lesssim 2 \Msun)$, which eventually explodes as a USSN. Indeed, some of the rapidly evolving transients such as SN 2005ek \citep{2013ApJ...774...58D}, iPTF14gqr \citep{2018Sci...362..201D}, and SN 2019dge \citep{2020ApJ...900...46Y} are suggested to be possible candidates for USSNe \citep{2017MNRAS.466.2085M}. In addition, it has been proposed that during the operation period of the Zwicky Transient Facility \citep[ZTF,][]{2019PASP..131g8001G, 2019PASP..131a8002B}, roughly $10$ USSNe within $300\,$Mpc will be detected per a year \citep{2019ApJ...882...93H}. Hence, it is expected that future surveys and follow-up observations of transients will enable us to examine in detail the validity of the USSN scenario as the formation mechanism of DNS binaries.

Another way to experimentally test the USSN scenario is to search for supernova remnants (SNRs) hosting a DNS binary. After the explosion, the ejecta of the USSN sweeps up the surrounding CSM while expanding into the interstellar space. Intriguingly, this kind of system can be potentially detected as a SNR hosting a DNS binary, which we will refer to as an ultra-stripped supernova remnant (USSNR) hereafter. While the current SNR surveys have not identified any of these remnants so far, we note that the observable characteristics of a USSNR have not been discussed and quantified in the literature. It is hence essential to investigate the dynamical evolution and emission properties of USSNRs using a dedicated simulation model to shed light on how they can be identified.

\cite{2013ApJ...778L..23T} developed a progenitor evolution model for the USSN, and showed that the mass-transfer rate through RLO can be enhanced up to $\dot{M} \sim 10^{-5} \Msunperyear$ in the last 0.1 Myr prior to the core collapse. Because the mass-transfer rate is orders of magnitude larger than the Eddington accretion rate onto the neutron star, a large fraction of the stripped gas escapes the binary system and distributes around the progenitor as CSM. Assuming a wind velocity $v_w \sim 1000\,{\rm km}\,{\rm s}^{-1}$, the gas which has been expelled from the binary system in the RLO phase can reach a distance of $\sim 100\,{\rm pc}$ from the progenitor, implying that the evolution of the USSNR is heavily influenced by the CSM created by the RLO mass loading process. However, detailed models for the mass-loss history driven by binary interaction are in most cases not incorporated in the simulations of SNR dynamics, which is particularly critical for understanding the properties of USSNRs.

In this study, we investigate the characteristics of a USSNR using a grid of one-dimensional hydrodynamic simulations. By employing the binary evolution model presented in \cite{2013ApJ...778L..23T}, we first construct the large-scale structure of the CSM surrounding the USSN progenitor. We next calculate the hydrodynamics of the USSN ejecta interacting with the composed CSM and the resulted synchrotron radiation. Our simulations reveal that the blastwave of USSNRs has a difficulty in penetrating the hot plasma, which had been shaped by the preceding mass loss from the progenitor binary. Radio emission from a young USSNR is predicted to be bright enough to be detected if it inhabits our Galaxy, while its luminosity starts to decrease at $t\gtrsim 10^3\,$years, making the USSNR observable for a relatively short time period. Besides, the low surface brightness of a USSNR predicted by our models at its typical diameters ($D\sim \mathcal{O}(10\,$pc$)$) can serve as a key to the identification of these remnants in the future.

This paper is organized as follows. In Section~\ref{sec:progenitor}, we review the USSN scenario as a formation theory of DNS, and describe the progenitor models used in our simulations. In Section~\ref{sec:CSMformation}, we discuss the formation sequence of the CSM, followed by a description of the procedures for constructing our CSM models. In Section~\ref{sec:SNR}, we examine the hydrodynamic evolution of a USSNR and show the properties of the expected radio signals, including the light curve and surface brightness. Their implications are discussed in Section~\ref{sec:discussion}, and our results are summarized in Section~\ref{sec:summary}.

\section{Progenitor model}~\label{sec:progenitor}

\cite{2013ApJ...778L..23T} investigated the binary stellar evolution of a $2.9 \Msun$ He star with a neutron star companion, having an initial orbital period of $0.1$ day. They found that the He star reduces its own mass down to $1.5\Msun$ through RLO, and suggested that the He star explodes as a USSN which can be a candidate for some rapidly evolving transients. Here we overview the stellar evolution of the progenitor of a USSN, which is crucial for understanding the formation of the CSM adopted in this study.

\begin{figure*}[ht!]
\plotone{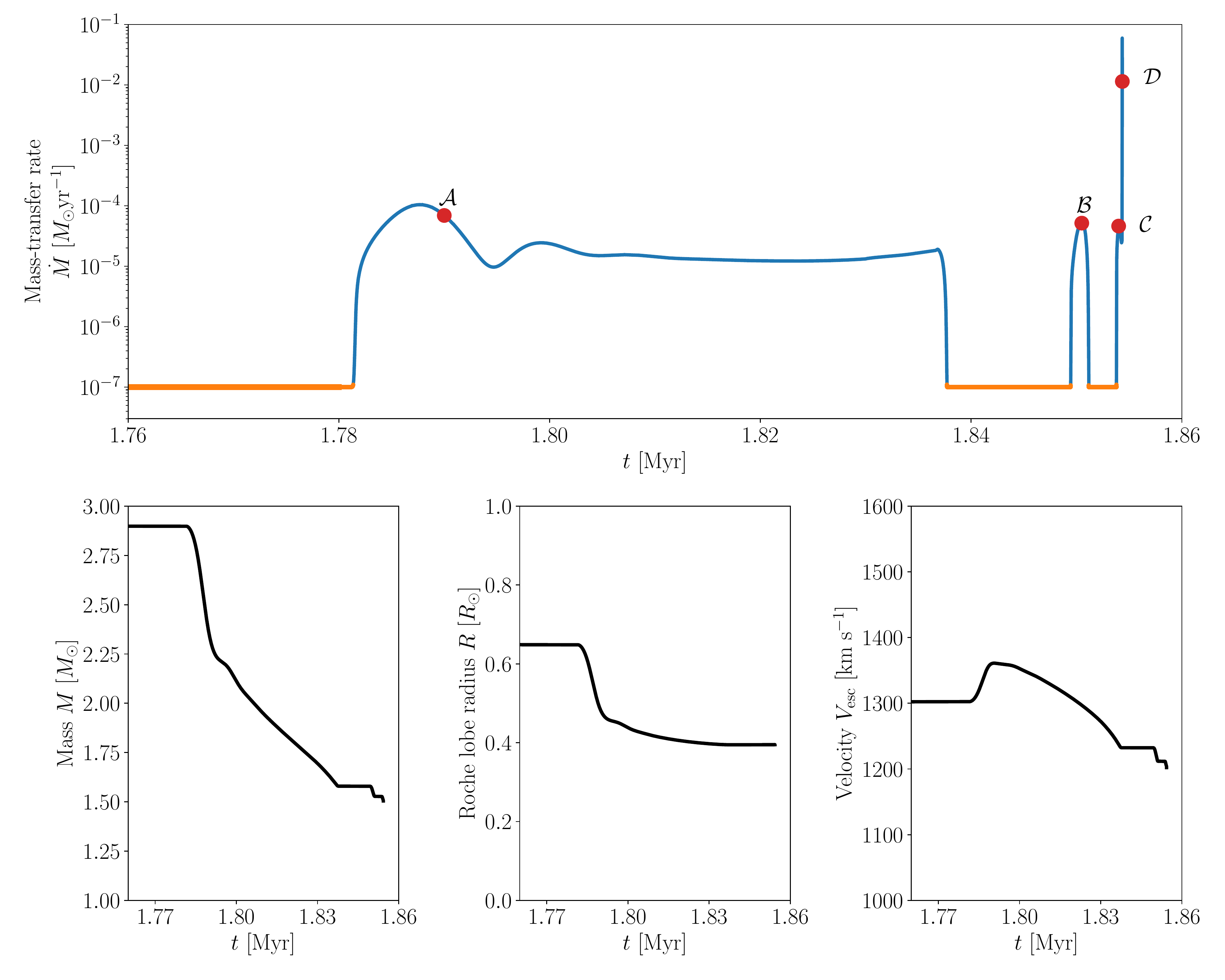}
\caption{Time evolution of the mass-loss history (top), total mass (bottom left), Roche lobe radius (bottom middle), and escape velocity (bottom right) of the USSN progenitor. Blue and orange lines correspond to the phases in which the binary system experiences RLO or not, respectively. Note that within the first $1.78$ Myrs the progenitor is in a stable core He-burning stage, so that these values are kept constant.}
\label{fig:progenitor}
\end{figure*}

Figure~\ref{fig:progenitor} shows the time evolution of the mass $(M)$, radius of the Roche lobe $(R)$, escape velocity $(V_{\rm esc})$, and mass-transfer rate $(\dot{M})$ of the USSN progenitor presented in \cite{2013ApJ...778L..23T}. Here, the escape velocity is defined as $V_{\rm esc} = \sqrt{2GM/R}$, where $G$ is the gravitational constant. When the progenitor is in the state illustrated by the blue line, its outer layer is stripped away by the companion neutron star through RLO. Until the core collapse, the He star experiences RLO three times; the first phase is at 1.78 Myr $\lesssim t \lesssim$ 1.84 Myr during which the core has exhausted its He-burning fuel ($\mathcal{A}$). The second is at $t\sim 1.851$ Myr when the core C-burning has ended ($\mathcal{B}$), and the third is at $t\gtrsim 1.854$ Myr in which the off-center O-burning is about to onset ($\mathcal{C}$ and $\mathcal{D}$). The CSM around the USSN progenitor is hence expected to be shaped by these three phases of mass loss activities. We note that the progenitor spends most of its lifetime in the state shown by the orange line prior to $\mathcal{A}$, and that the increase of the mass-loss rate is realized in the last $0.1$ Myr before core collapse. 

The progenitor does not experience RLO in the detached phases, during which we conservatively assume a mass loss rate of $10^{-7} \Msunperyear$. This mimics the stellar wind from the progenitor, but the mass and kinetic energy released by this wind are smaller than those carried by the gas stripped away through the RLO. Thus, we can assume that the stellar wind from the progenitor has an insignificant influence on the overall wind hydrodynamics, and that the consistency with the stellar evolution model is maintained. The model developed by \cite{2013ApJ...778L..23T} covers the lifetime of the He star only until $10\,$years prior to its core collapse. To trace the evolution up to the moment of the explosion, we use the final values of $M, R, V_{\rm esc}$, and $\dot{M}$ from the model for the last $10\,$years. 

The gas transferred from the progenitor first flows toward the neutron star with an accretion rate orders of magnitude larger than the Eddington accretion rate \citep{2013ApJ...778L..23T}. The neutron star cannot feed up anymore and thus drives the accreted gas outward by mechanisms such as propeller effect \citep{2017ApJ...846..170T}. However, resolving the detail of this outflow dynamics is beyond the scope of this work. For simplicity, we assume that the material which has been stripped away from the He star launches outward spherically at the radius of the Roche lobe $R$ with a velocity $V_{\rm esc}$ and mass-loss rate $\dot{M}$. Then, the mass density at the Roche lobe radius $({\dot{M}}/{4\pi R^2 V_{\rm esc}})$ can be estimated. Given the density and velocity of the gas at the Roche lobe radius as an inner boundary condition, we can solve the hydrodynamics of the gas launched from the progenitor binary to model the CSM formation around the progenitor. Combined with a parametric survey described in the following sections, this strategy allows us to demonstrate the long-term evolution properties of a USSNR with the mass loss history of the progenitor taken into account.

\section{CSM formation}\label{sec:CSMformation}
In this section, we describe our procedure for modeling the formation of the CSM surrounding the USSN progenitor. First, we construct the initial profile of the interstellar medium (ISM) in Section~\ref{sec:initial}. We then explain our methodology for simulating the hydrodynamics of the mass-loss material in Section~\ref{sec:CSMhydro}, and the properties of the composed CSM in Section~\ref{sec:CSMresult}.

\subsection{Initial setup}\label{sec:initial}

The progenitor experiences a hydrogen-rich envelope ejection driven by the CE interaction before the stripping of the helium gas through RLO. The distribution of this expelled hydrogen-rich gas is important because it interacts with the helium gas released through RLO later on. Although {some recent multi-dimensional simulations have succeeded in completely ejecting the hydrogen-rich envelope of a red supergiant through the CE interaction under some assumptions and realizations (\citealp[but see also]{2020arXiv201106630L, 2021arXiv211100923L} \citealp[][]{2021arXiv210714526V})}, the distribution of the material ejected by the CE interaction is still not completely understood.

\begin{figure}[ht!]
\plotone{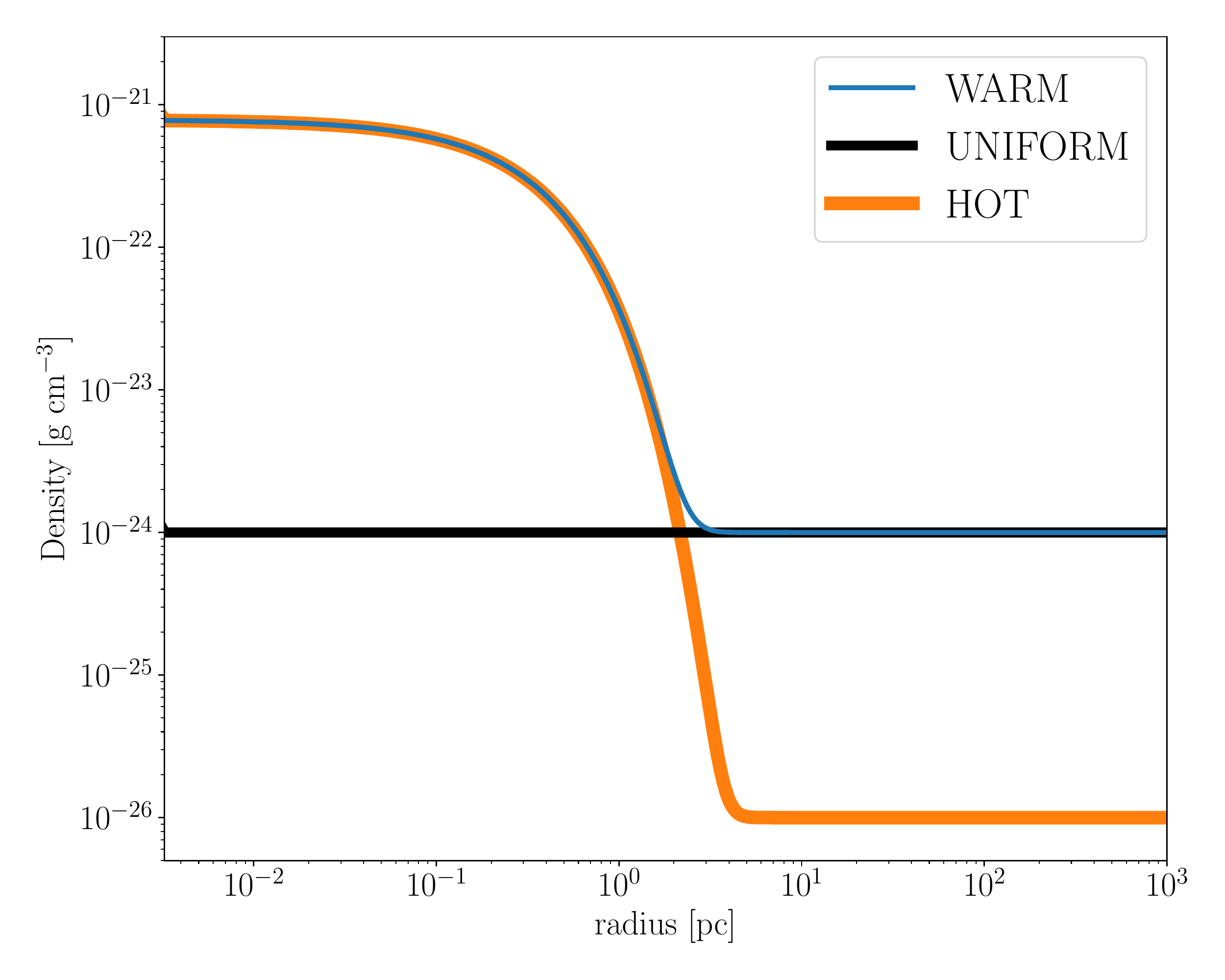}
\caption{Initial density profiles of the ISM with a component from CE ejection. {Blue, orange, and black lines represent the model `WARM', `HOT' and `UNIFORM'}, respectively.}
\label{fig:initialISM}
\end{figure}

Figure~\ref{fig:initialISM} shows {three} models we adopt for the initial density profile of the CE material. We consider a situation where the ejected gas with a mass $M_{\rm CE} = 10\Msun$ is distributed within a radius $R_{\rm CE}$ which smoothly connects with the ISM. Given that the characteristic timescale of the CE interaction is around thousands of years \citep{2013AARv..21...59I}, the gas ejected with a speed $\sim 100$ km s\pow{-1} can reach a radius $R_{\rm CE} \sim 10^{18}$ cm. {Since there is a variety in the ISM properties such as density and temperature \citep[e.g.,][]{2008MNRAS.390L..19B,2011piim.book.....D}, we consider two ISM phases; a warm phase ($\rho_{\rm ism}=10^{-24}$\,g\,cm\pow{-3}, $T_{\rm ism}=10^4$\,K) and a hot phase ($\rho_{\rm ism}=10^{-26}$\,g\,cm\pow{-3}, $T_{\rm ism}=10^6$\,K).} {We remark that the thermal pressure in these two initial profiles are equal to each other.} In addition, we prepare a reference model `UNIFORM', in which a static and uniform ISM resides throughout the simulation domain with a density $\rho_{\rm ism}=10^{-24}$\,g\,cm\pow{-3}, to {evaluate} the effect \deleted{of the absence} of the CE ejection activity. The specific profiles of the initial density for each model are described {in Table \ref{tab:profiles}.} The derivation of the exact value of $\rho_{\rm CE}$ is explicated in Appendix~\ref{sec:app_rhoce}. We consider a static ISM profile ($v=0$). The initial velocity profile of the CE component does not have an important role in the hydrodynamics of the CSM formation because the expected $V_{\rm CE}$ is negligibly lower than the velocity of the wind from the progenitor binary. {To verify this we conducted simulations in which the initial velocity of the CE component is {assumed to be $100\,$km\,s\pow{-1}} and confirmed that the outcome is not changed.} We assume the temperature $T=T_{\rm ism}$ {and a solar metallicity throughout the entire profiles at this stage.} A comparison of the results among these models enables us to evaluate how much the properties of the CE ejection affect the CSM formation and the subsequent SNR evolution.

\begin{deluxetable}{cccc}[ht!]\label{tab:profiles}
\tablecaption{Parameters for initial profiles}
\tablehead{
\colhead{name} &
\colhead{$\rho_{\rm ism}$\,[g\,cm\pow{-3}]} &
\colhead{$\rho(r)$} &
\colhead{$T_{\rm ism}$\,[K]}
}
\startdata
WARM & $10^{-24}$ & $\rho_{\rm CE}\exp(-r/R_{\rm CE}) + \rho_{\rm ism}$ & $10^4$ \\
HOT & $10^{-26}$ & $\rho_{\rm CE}\exp(-r/R_{\rm CE}) + \rho_{\rm ism}$ & $10^6$ \\
UNIFORM & $10^{-24}$ & $\rho_{\rm ism}$ & $10^4$
\enddata
\end{deluxetable}

\subsection{Wind hydrodynamics}\label{sec:CSMhydro}
We solve the one-dimensional equations of ideal gas hydrodynamics where the internal energy is taken away by radiative cooling in spherical coordinates. The governing equations are described as follows:
\begin{eqnarray}
    &\partialdiff{\rho}{t} + \frac{1}{r^2}\partialdiff{(r^2\rho v)}{r} = 0, \\
    &\partialdiff{v}{t} + v\partialdiff{v}{r} = -\frac{1}{\rho}\partialdiff{p}{r}, \\
    &\partialdiff{}{t}\left[\rho\left(\frac{1}{2}v^2 + e\right)\right] + \frac{1}{r^2}\partialdiff{}{r}\left[r^2\rho v\left(\frac{1}{2}v^2 + h\right)\right] \nonumber \\
    & = -n_i n_e\Lambda(T),
\end{eqnarray}
where $\rho$ is the mass density, $v$ is the velocity,
$p$ is the pressure, $e$ is the specific internal energy, $h = e+p/\rho$ is the specific enthalpy, $n_i$ and $n_e$ are the number density of ions and electrons. $\Lambda(T)$ represents the radiative cooling function, for which we employ the power-law formalism introduced by \cite{1994ApJ...420..268C}. {The energy loss by radiative cooling is calculated only in the optically thin region where $\tau\leq 1$ \deleted{after the shock breakout } which is sufficient for tracing the evolution of the blastwave (see also Section \ref{sec:radiativecooling}).} These governing equations are closed with the equation of state, $p = (\gamma-1) \rho e$, where $\gamma = 5/3$ is the adiabatic index. {The equations are solved by a Roe Riemann solver with the second entropy fix by Harten and Hyman to treat the contact discontinuity and the shock wave \citep{1983JCoPh..50..235H}. }The numerical accuracy of the code used in this study is verified in Appendix~\ref{sec:numerics}.

We divide the simulation domain from $10^{16}\,{\rm cm}$ to $3\times 10^{21}\,{\rm cm}$ into 2047 zones in a logarithmic scale. Inside $10^{16}\,{\rm cm}$ as an inner boundary condition, we inject the blowing He-rich gas whose time evolution is described in Section~\ref{sec:progenitor}. 

We trace the distribution of the chemical abundances by advection, assuming that no mixing of the chemical composition occurs. The abundance distribution is required in order to accurately estimate the number density of ions and electrons in the radiative cooling term.

\subsection{Composed CSM}\label{sec:CSMresult}

\begin{figure}[ht!]
\plotone{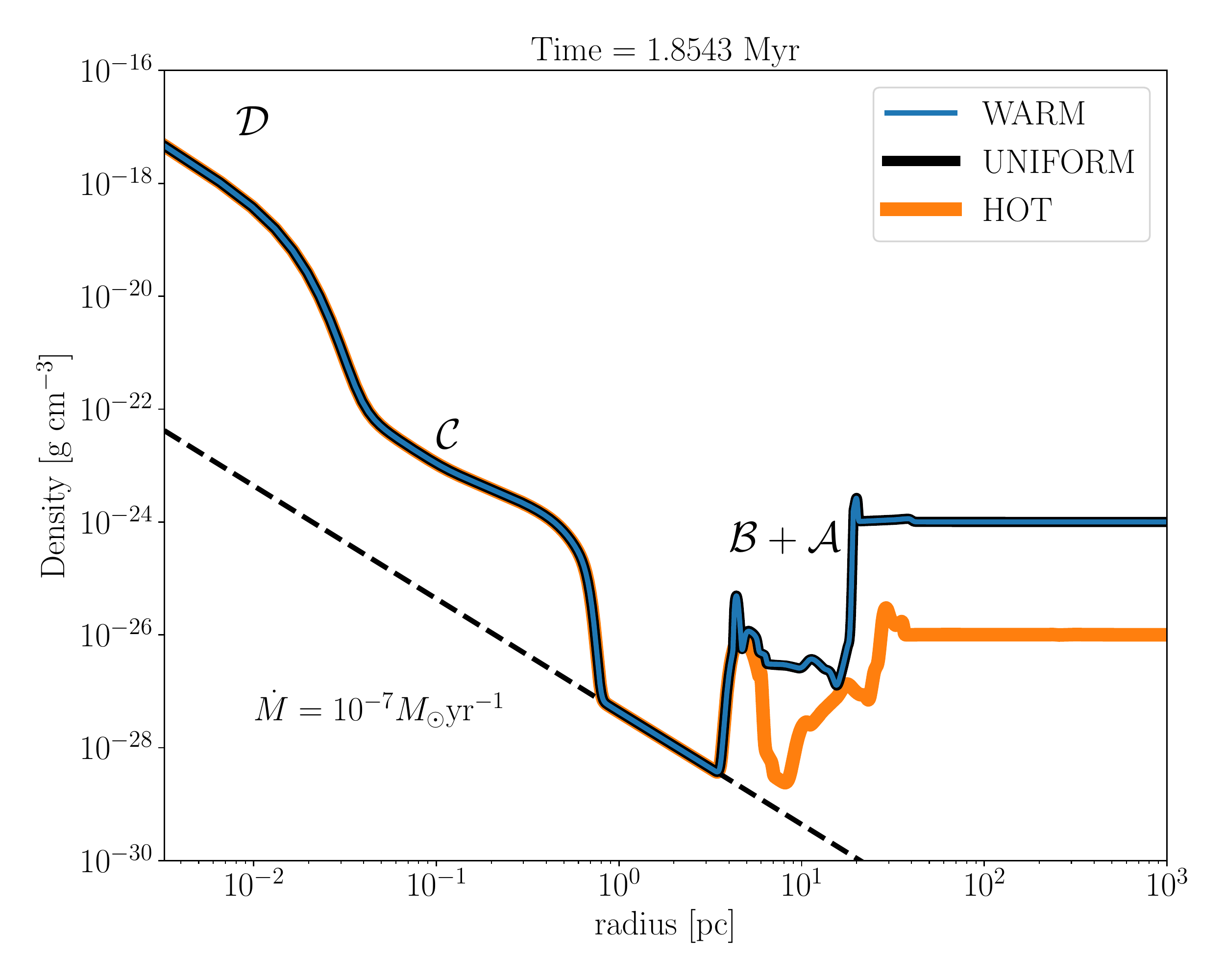}
\caption{The density structure of the composed CSM. The dashed black line shows the distribution realized for the steady wind with its mass-loss rate $\dot{M}=10^{-7}\,\Msunperyear$. The distributions of the gas pointed out by cursive alphabets represent that they are from the mass loss activity referred in Figure~\ref{fig:progenitor}.}
\label{fig:CSMmodel_rho}
\end{figure}

Figure~\ref{fig:CSMmodel_rho} shows the snapshot of the density structure of the CSM at the moment of core collapse of the progenitor. {The model `WARM' and `UNIFORM' share an identical CSM structure in the entire simulation domain. It is also the case for the model `HOT' within $\sim 3\,$pc, but its outer configuration deviates from the other two models.} The distribution of the density within $\sim 3\,$pc reflects the mass loss history. Namely, the dense CSM being distributed around $r\sim0.01\,{\rm pc}$ and $0.1\,{\rm pc}$ are originated from the mass loss at points $\mathcal{D}$ and $\mathcal{C}$ in Figure~\ref{fig:progenitor}. Yet, a segment resides around $10\,{\rm pc}$ in which the density is roughly constant with some fluctuations. This non-smooth segment is created by the collision between the wind launched at point $\mathcal{B}$ and the reverse shock generated by the gas ejected at point $\mathcal{A}$ before. The ISM wall is located at a radius of $20\,$pc {in the model `WARM' and `UNIFORM' and $30\,$pc in the model `HOT', respectively.}

We will briefly elaborate on the importance of the CE component on the ISM profile. The reference model `UNIFORM' without the CE component allows us to investigate the contribution of the CE component on the hydrodynamics of the wind. The results obtained from this reference model are found to be almost identical to the outcome from `WARM', being nearly indistinguishable just in Figure~\ref{fig:CSMmodel_rho}. This can be interpreted as follows. {The radius of the ISM wall} is roughly determined by the balance between the ram pressure of the wind and the thermal pressure of the swept-up material \citep{1977ApJ...218..377W}, which is computed as $\sim 20\,{\rm pc}$ in our simulations. The enclosed mass of the initial ISM profile at $r\sim 20\,{\rm pc}$ is $\sim 400M_\odot$, indicating that the mass of the CE component can be regarded to be negligibly small. Hence, the composed CSM has similar characteristics between `WARM' and `UNIFORM'. {We confirmed that even when considering a uniformly distributed hot ISM {($\rho_{\rm ism}=10^{-26}\,{\rm g\,cm}^{-3},\,T{\rm ism}=10^6\,$K)}, the consequent CSM structure does not differ from the model `HOT' significantly other than slight quantitative modifications.} This implies that as long as the CE ejection before the USSN is considered within a range of typical time and energy scales, it does not play an important role in the formation of the CSM around the USSN progenitor.

\begin{figure}[ht!]
\plotone{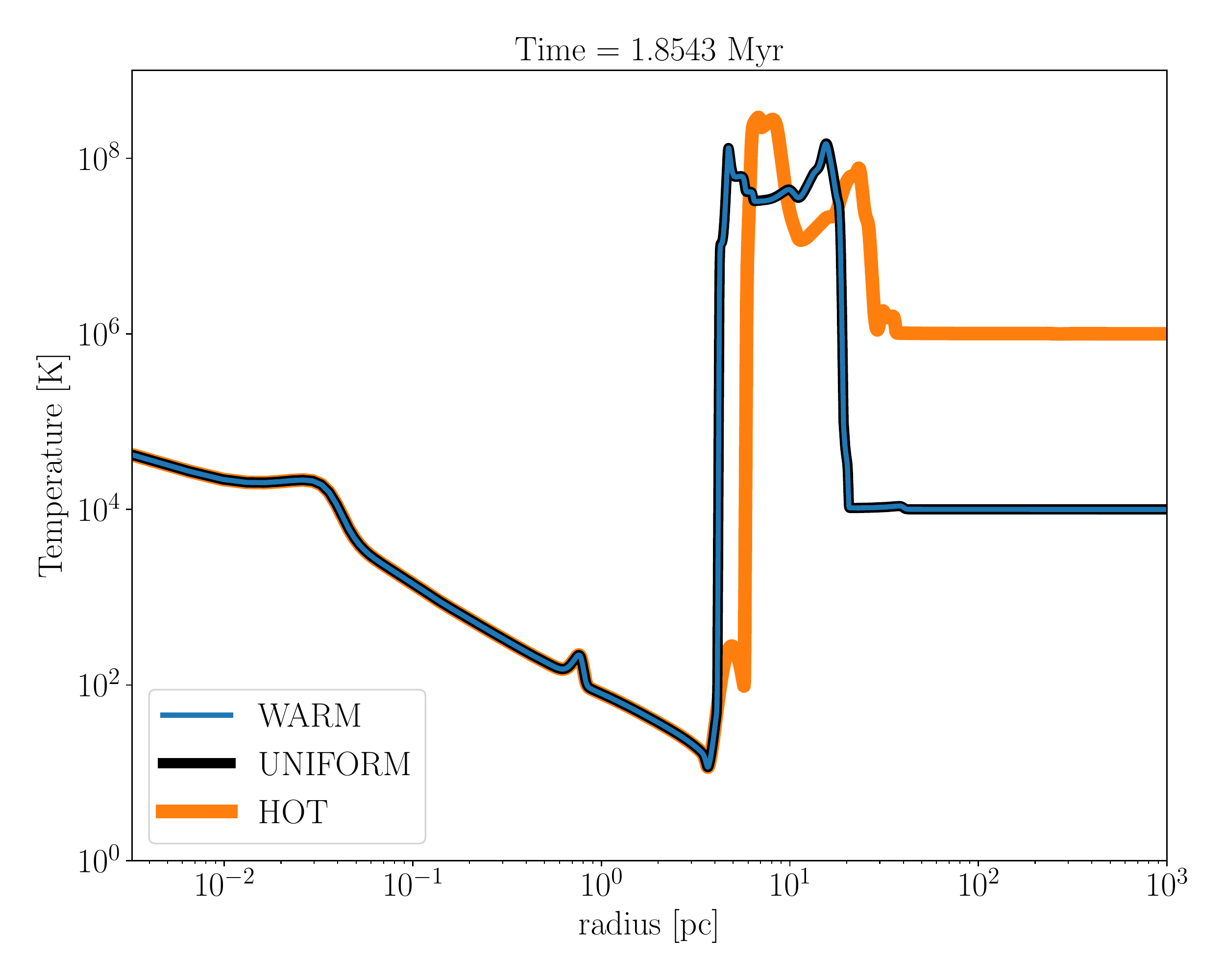}
\caption{The temperature structure of the composed CSM.}
\label{fig:CSMmodel_temp}
\end{figure}

Figure~\ref{fig:CSMmodel_temp} shows the temperature structure of the CSM at the moment of core collapse. Similar to the density structure, {the models `WARM' and `UNIFORM' have the same temperature structure over the entire region. The model `HOT' also possesses the identical distribution with the other models within $3\,$pc, but the quantitatively different structure is formed outside  $3\,$pc}. A hot plasma {with $\sim 10^8\,$K} is located in the vicinity of the ISM wall in {all} models. The location of the inner edge of this hot plasma coincides with the radius of the termination shock of the wind driven by RLO. The geometrical thickness of the plasma is $\sim 10\,$pc. The existence of this hot plasma region plays a critical role in weakening the SNR blastwave as it propagates through the region as discussed later.

\section{SNR evolution} \label{sec:SNR}
In this section, we investigate the evolution of a USSNR interacting with the CSM constructed in the previous section. In Section~\ref{sec:SNRmethod}, we show the method to simulate the dynamics of the ejecta and the expected synchrotron emission, and the results are presented in Section~\ref{sec:USSNRresult}. As was confirmed in the previous section, the solution derived from the model without a CE component (`UNIFORM') converges to that of `WARM'. We will therefore examine results from the {models} `WARM' and `HOT' hereafter.

\subsection{Method}\label{sec:SNRmethod}

\subsubsection{Ejecta dynamics}\label{sec:ejectahydro}
The initial profile of the USSN progenitor is taken from \cite{2017MNRAS.466.2085M}, who evolved the model of the He star previously presented by \cite{2013ApJ...778L..23T} further until core collapse. Then we attach the CSM composed in Section~\ref{sec:CSMformation} to the progenitor while retaining the distribution of the density, velocity, temperature, and chemical abundance. 

We next examine the hydrodynamics of the SN explosion to obtain the SN ejecta structure. We excise the remnant mass $M_{\rm rem}=1.35\Msun$ from the inner region of the progenitor, and inject an explosion energy $E_{\rm exp} = 10^{50}$\,erg to the rest of the material in the progenitor ($M_{\rm ej} \sim 0.15\Msun$) {as a thermal energy} following the method developed by \cite{2015ApJ...814...63M}. The explosion energy is chosen based on light curve models \citep{2017MNRAS.466.2085M}, which is also consistent with that proposed by state-of-the-art simulations \citep{2015MNRAS.454.3073S, 2018MNRAS.479.3675M}. The profile is resolved into more than 4000 meshes with a logarithmic spacing, and the hydrodynamics of the ejecta is calculated by the same method as described in Section~\ref{sec:CSMhydro}, except that a reflective condition is employed at the inner boundary. As a result, we obtain the time evolution of the blastwave velocity and the trajectory of Lagrangian particles, which are used to compute the energy distribution of relativistic electrons and the amplified magnetic field (see the next section).

As the SNR evolves into the Sedov phase, its reverse shock begins to propagate towards the inner region and heats up the ejecta \citep[e.g.,][]{1999ApJS..120..299T}. Since the simulation domain is resolved under a logarithmic mesh, the high temperature in the inner region can cause small timesteps, making it difficult for the simulation to progress. To solve this numerical difficulty, we excise the Eulerian meshes in the innermost region within $10^{18}$ cm when the blastwave radius has reached $10^{19}$ cm. This does not affect the consistency of the simulations since the total gas mass within $10^{18}$ cm at the moment of the excision is negligibly small and hence dynamically unimportant. This allows us to trace the long-term evolution of the USSNR within a reasonable simulation time. The computations are terminated at $10^5\,$years since the explosion.

\subsubsection{Particle acceleration and magnetic field amplification}\label{sec:PAMFA}

Once the gas is heated by the forward shock, the diffusive shock acceleration (DSA) imparts relativistic energies to the injected charge particles and induces amplification of the turbulent magnetic field \citep[e.g.,][]{1949PhRv...75.1169F, 1983RPPh...46..973D}. The region shocked by the blastwave serves as a site of synchrotron emission from SNRs \citep{2008ARAA..46...89R, 2015AARv..23....3D}. In this study, we define the blastwave {as the discontinuity which satisfies the following two conditions: (1) the pressure jump is the largest in the simulation domain, and (2) the Mach number is greater than 3. The latter is justified because strong shocks have a potential to drive DSA, whilst weak shocks are less capable of efficient particle acceleration, confirmed by the observations for radio relics in galaxy clusters \citep[e.g.,][and references therein]{2020AA...634A..64B}.}

We first consider a Lagrangian mesh $a_s$ through which the blastwave passes at time $t_s$. As the shock sweeps through the mesh, the charged particles are accelerated to relativistic energies, coupled with an amplification of the turbulent magnetic field. We model the energy densities of the accelerated relativistic electrons $(u_e)$ and the magnetic field $(u_B)$ in the Lagrangian mesh $a_s$ as follows:
\begin{eqnarray}
    && u_e = \epsilon_e \rho_{\rm sh} (V_b - v_u)^2, \\
    && u_B = \frac{B^2}{8\pi} = \epsilon_B \rho_{\rm sh} (V_b - v_u)^2,
\end{eqnarray}
where $\epsilon_e$ and $\epsilon_B$ are the acceleration and amplification efficiencies, $\rho_{\rm sh}$ is the mass density in the Lagrangian mesh $a_s$, $V_b$ is the velocity of the blastwave, and $v_u$ is the velocity of the unshocked gas upstream of the shock, respectively. These parametrizations are conventionally used in the modeling of radio SNe \citep[e.g.,][]{2006ApJ...641.1029C, 2006ApJ...651..381C, 2019ApJ...885...41M}. These equations apply to the mesh only when $V_b - v_u>0$.

{The energy distribution of the accelerated electrons, $N(a_s, E)$, }is described by a power-law distribution as follows:
\begin{eqnarray}
    N(a_s, E) = C E^{-p},
\end{eqnarray}
where $E$ and $p$ are the energy and the spectral index of the electrons, respectively. The coefficient $C$ is determined by performing a normalization of the energy density:
\begin{eqnarray}~\label{eq:NEnorm}
    \int_{E_{\rm min}}^\infty E N(a_s, E) dE = u_e,
\end{eqnarray}
{where $E_{\rm min}=2m_e c^2$ is used (see Section~\ref{sec:ee_Emin} for a discussion on the uncertainty related to $E_{\rm min}$).}

As the system evolves, the ejecta expands and the blastwave propagates to the next Lagrangian mesh. Meanwhile, the relativistic electrons lose their energies by both synchrotron and adiabatic cooling, and the magnetic field also decays with the adiabatic expansion. We consider a Lagrangian mesh $(a)$ which had been heated by the shock at mass coordinate $a_s$ and time $t_s$, and assume that the relativistic particles are confined within the mesh and the magnetic field is frozen in the plasma. We calculate the cooling processes of the accelerated particles and the time evolution of the energy distribution following previous studies \citep[e.g.,][]{1998ApJ...493..375R,2011AA...526A.129O,2014ApJ...789...49F}. An electron's energy $E$ declines to $E'$ through synchrotron and adiabatic cooling, which can be written as follows:
\begin{eqnarray}
	&&E'(a,t) = E {\displaystyle \frac{\alpha(a,t)^{\frac{1}{3}}}{1+\Theta(a,t) \frac{E}{m_e c^2}}}, \\
	&&\alpha(a,t) = {\displaystyle \frac{\rho(a,t)}{\rho(a_s, t_s)}}, \\
	&&\Theta(a,t) = {\displaystyle \int_{t_s}^t \frac{\lambda}{m_e c^2}B(a,t')^2\alpha(a,t')^{1/3} dt'}, \\
 	&&\lambda = {\displaystyle \frac{4q^4}{9m_e^2 c^3}}
\end{eqnarray}
where $c$ is the speed of light, $q$ is the elementary charge, and $m_e$ is the mass of electron, respectively. The energy distribution of the electrons evolves following number conservation, i.e.
\begin{eqnarray}
    N(a,E') = N(a_s, E)\frac{dE}{dE'}.
\end{eqnarray}
As for the strength of the magnetic field, we consider a magnetic flux conservation in each Lagrangian mesh.

\subsubsection{Synchrotron emission}\label{sec:synchrotron}

Given the energy distribution of electrons and the strength of the magnetic field, the intensity of the synchrotron emission $I_\nu$ can be calculated by integrating the radiative transfer equation written as follows:
\begin{eqnarray}
    \normaldiff{I_\nu}{a} = 
    \left \{
    \begin{array}{ll}
            -\alpha_{\nu, {\rm syn}} I_\nu + j_{\nu, {\rm syn}} & (a \leq R_b) \\
            -\alpha_{\nu, {\rm ff}} I_\nu & (a > R_b) 
    \end{array}
    \right.
\end{eqnarray}
where $j_{\nu, {\rm syn}}, \alpha_{\nu, {\rm syn}}$, and $\alpha_{\nu, {\rm ff}}$ are the synchrotron emissivity, synchrotron self-absorption and free-free absorption coefficient, respectively \citep{1979rpa..book.....R}, and $R_b$ is the blastwave radius. 

We also calculate the surface brightness $\Sigma(\theta)$ which is often used as a diagnostic observable for SNRs. {$\Sigma$-$D$ diagrams which show the relation between the surface brightness and the diameter of SNRs are} commonly used for determining the distance to the objects \citep[see e.g.,][and references therein]{1968AJ.....73...65P, 2013ApJS..204....4P}. Since the surface brightness $\Sigma(\theta)$ is independent of the distance to the SNR, it can be a useful quantity for investigating the intrinsic nature of the USSNR compared to the rest of the SNR population. $\Sigma(\theta)\delta \theta$, the power per unit surface area and unit frequency emitted from a ring with sky projection angles $\theta$ to $\theta+\delta \theta$, can be evaluated by integrating the total power of the synchrotron emission per unit volume along the line of sight as follows:
\begin{eqnarray}
    \Sigma(\theta) \delta \theta &=& \frac{\int_{\rm LOS} dl \epsilon_{\nu} \delta A(\theta)}{4\pi d^2 \Delta \Omega(\theta)} \simeq \int_{\rm LOS} dl \frac{\epsilon_{\nu}}{4} \frac{\delta \theta^2}{\theta^2}
\end{eqnarray}
where $d, \epsilon_{\nu} = 4\pi j_{\nu ,{\rm syn}}, \delta A(\theta) = \delta (\pi d^2 \theta^2)$, and $\Delta \Omega(\theta)$ are the distance to the SNR, the total power of the synchrotron emission per unit volume, the area of the ring with projection angle $\theta$, and the total solid angle of the SNR. The angle-averaged surface brightness can then be estimated, which allows us to examine the position of USSNRs on the $\Sigma$-$D$ diagram.

\subsubsection{Parameter sets}
Our treatment of DSA involves uncertainties from the DSA parameters $p, \epsilon_e$, and $\epsilon_B$. Although these parameters should in principle be constrained by particle-in-cell simulations \citep[e.g.,][]{PhysRevLett.114.085003, 2015ApJ...798L..28C}, the appropriate values are still debated. To investigate the dependence of the shock acceleration parameters on the radio light curves, we prepare 6 combinations of parameters chosen as follows. For the spectral index of electrons $p=2.1$ (hard), $p=2.5$ (intermediate), and $p=3.0$ (soft) are employed, while for the efficiency of particle acceleration and magnetic field amplification, the combinations $(\epsilon_e, \epsilon_B) = (10^{-2}, 10^{-1})$ \citep[typical for radio SNe,][]{2006ApJ...651..381C,2012ApJ...758...81M} and $(\epsilon_e, \epsilon_B) = (10^{-3}, 10^{-2})$ \citep[typical for SNRs,][]{2012ApJ...750..156L} are adopted. Our grid of 6 models for the shock acceleration parameters is then applied to the two kinds of CSM model `WARM' and `HOT'. The models are named by a {sequence of labels from the CSM model (`WARM' and `HOT')}, the first character of the word representing the spectral state (Hard, Intermediate, and Soft), and the object type (SN and SNR) for which the chosen value of the shock acceleration efficiency is typical. The different combinations of the CSM models and shock acceleration parameters are summarized in Table~\ref{tab:parameters}. {While our study employs time-independent values for the microphysics parameters $\epsilon_e, \epsilon_B,$ and $p$, it is possible that they vary with time depending on the hydrodynamic evolution of the shock front {(see Section~\ref{sec:ee_Emin} for a more detailed discussion)}.}

\begin{deluxetable}{ccccc}[ht!]\label{tab:parameters}
\tablecaption{Grid of models}
\tablehead{
\colhead{ID} &
\colhead{name} &
\colhead{$p$} &
\colhead{$(\epsilon_e, \epsilon_B)$}
}
\startdata
1 & WARM\_H\_SN & 2.1 & $(10^{-2}, 10^{-1})$ \\
2 & WARM\_H\_SNR & 2.1 & $(10^{-3}, 10^{-2})$ \\
3 & WARM\_I\_SN & 2.5 & $(10^{-2}, 10^{-1})$ \\
4 & WARM\_I\_SNR & 2.5 & $(10^{-3}, 10^{-2})$ \\
5 & WARM\_S\_SN & 3.0 & $(10^{-2}, 10^{-1})$ \\
6 & WARM\_S\_SNR & 3.0 & $(10^{-3}, 10^{-2})$ \\
7 & HOT\_H\_SN & 2.1 & $(10^{-2}, 10^{-1})$ \\
8 & HOT\_H\_SNR & 2.1 & $(10^{-3}, 10^{-2})$ \\
9 & HOT\_I\_SN & 2.5 & $(10^{-2}, 10^{-1})$ \\
10 & HOT\_I\_SNR & 2.5 & $(10^{-3}, 10^{-2})$ \\
11 & HOT\_S\_SN & 3.0 & $(10^{-2}, 10^{-1})$ \\
12 & HOT\_S\_SNR & 3.0 & $(10^{-3}, 10^{-2})$ \\
\enddata
\end{deluxetable}

\subsection{Characteristics of a USSNR}\label{sec:USSNRresult}
\begin{figure*}[ht!]
\centering
\includegraphics[width=0.9\columnwidth]{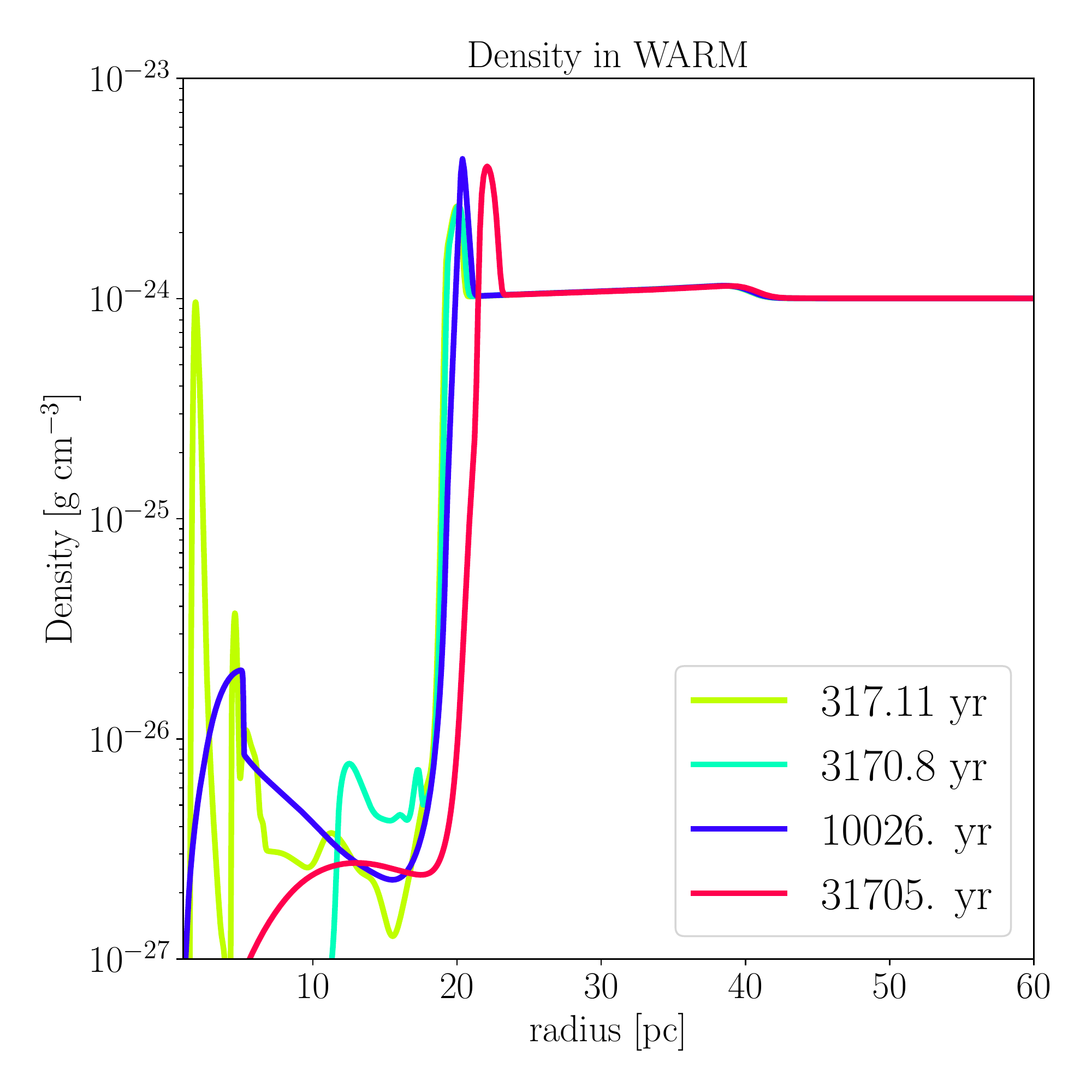}
\includegraphics[width=0.9\columnwidth]{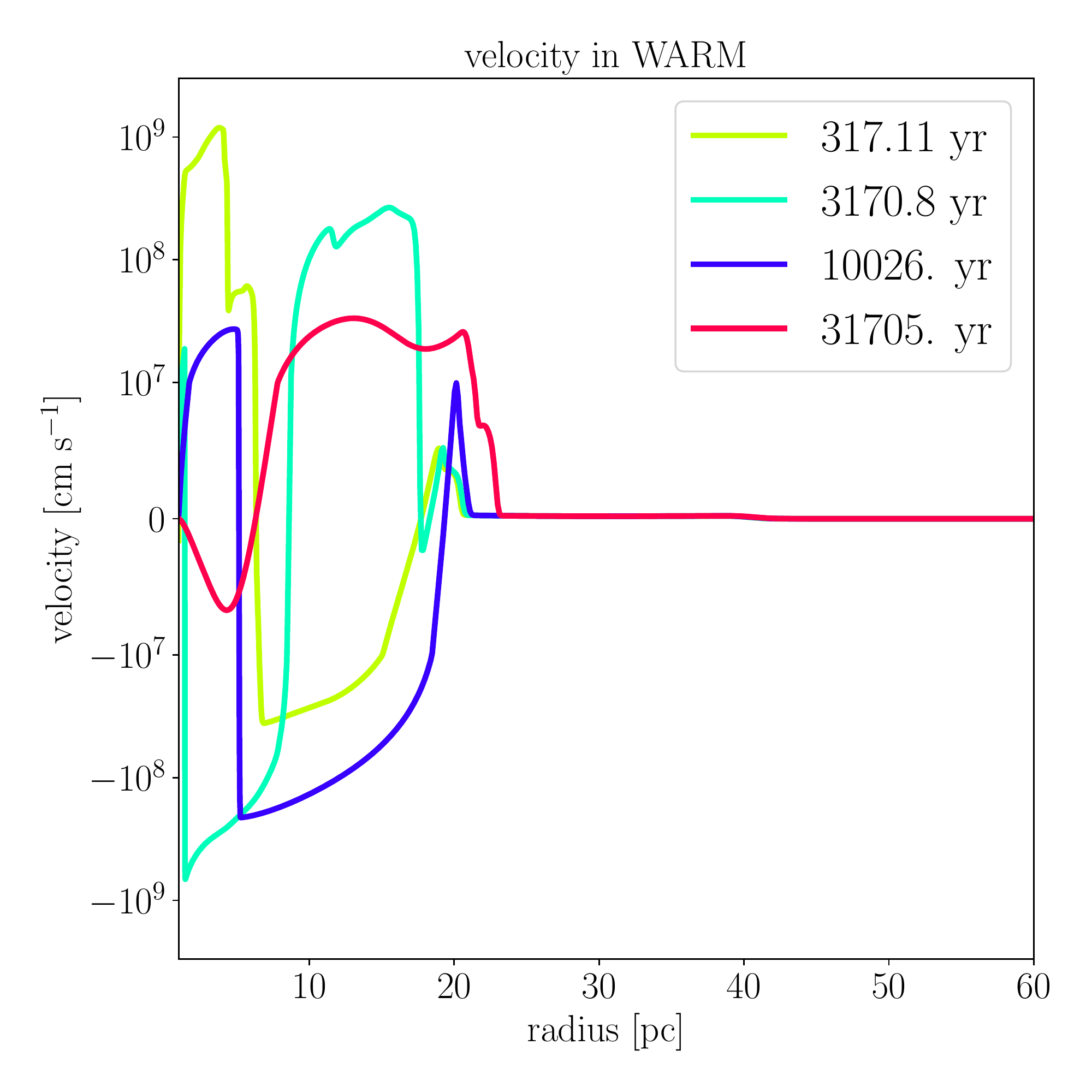}\\
\includegraphics[width=0.9\columnwidth]{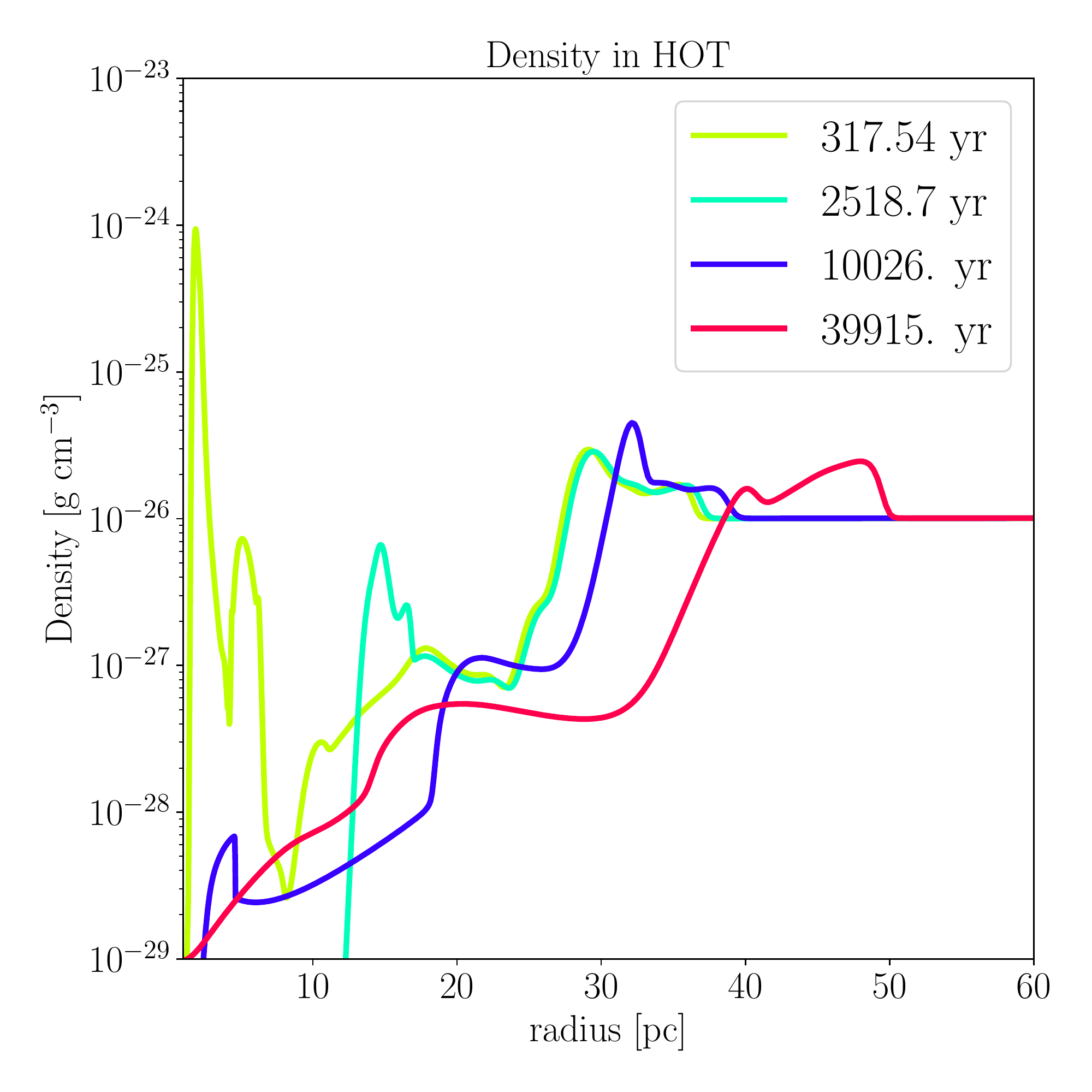}
\includegraphics[width=0.9\columnwidth]{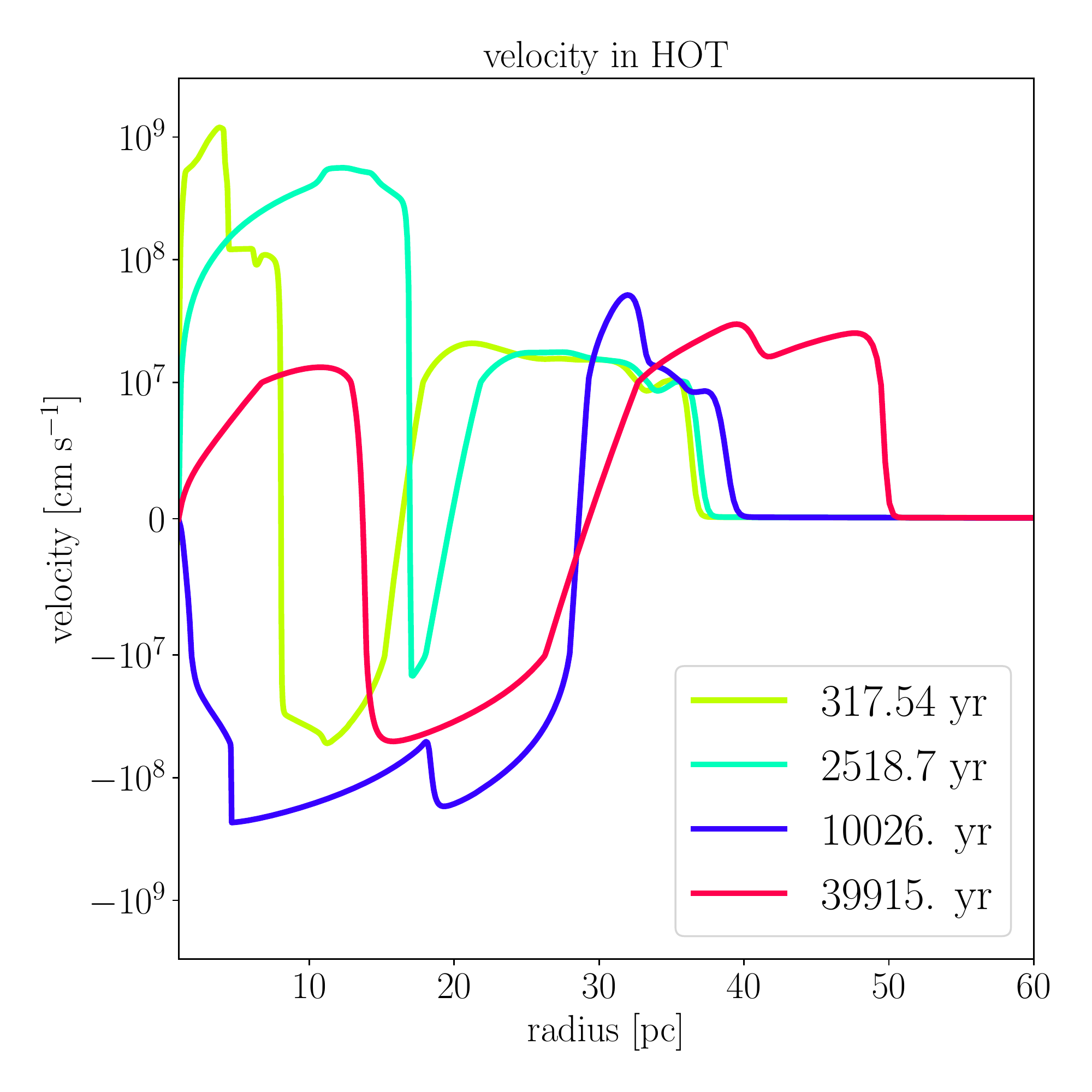}\\
\caption{Time evolution of the radial profiles of the density (left) and velocity (right) in `WARM' (top) and `HOT' (bottom).}
\label{fig:rho_vel}
\end{figure*}

\begin{figure*}
\centering
\includegraphics[width=0.9\columnwidth]{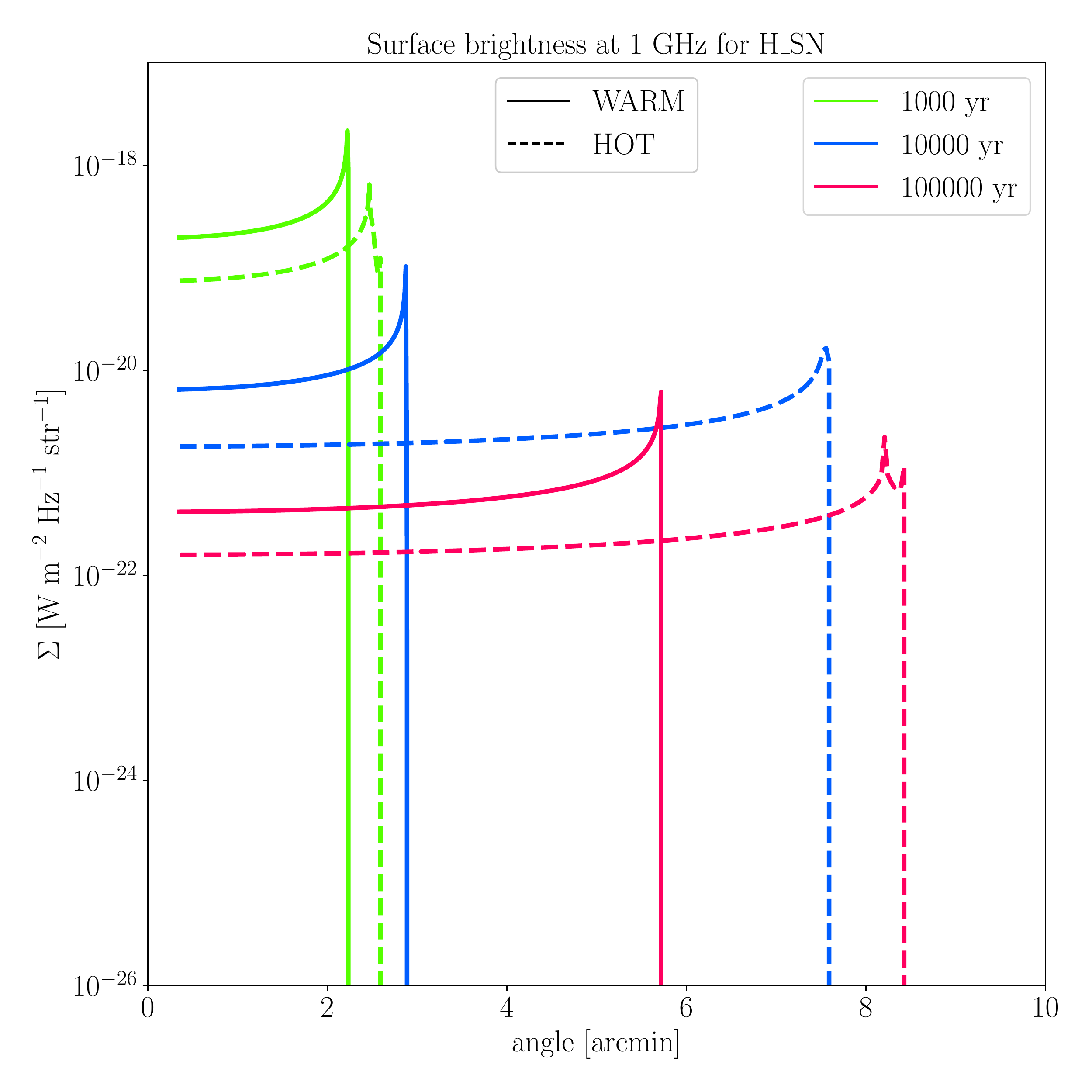}
\includegraphics[width=0.9\columnwidth]{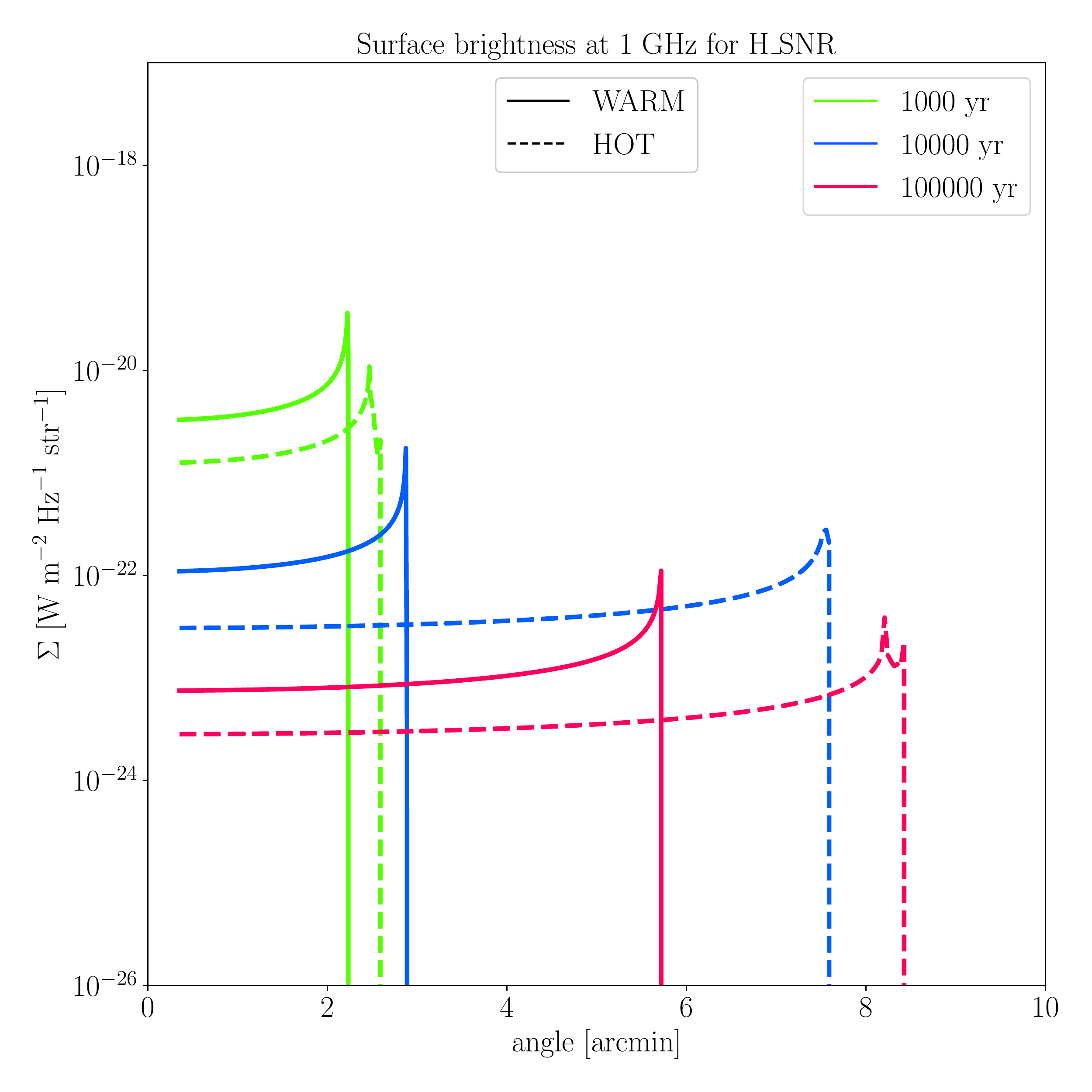}\\
\includegraphics[width=0.9\columnwidth]{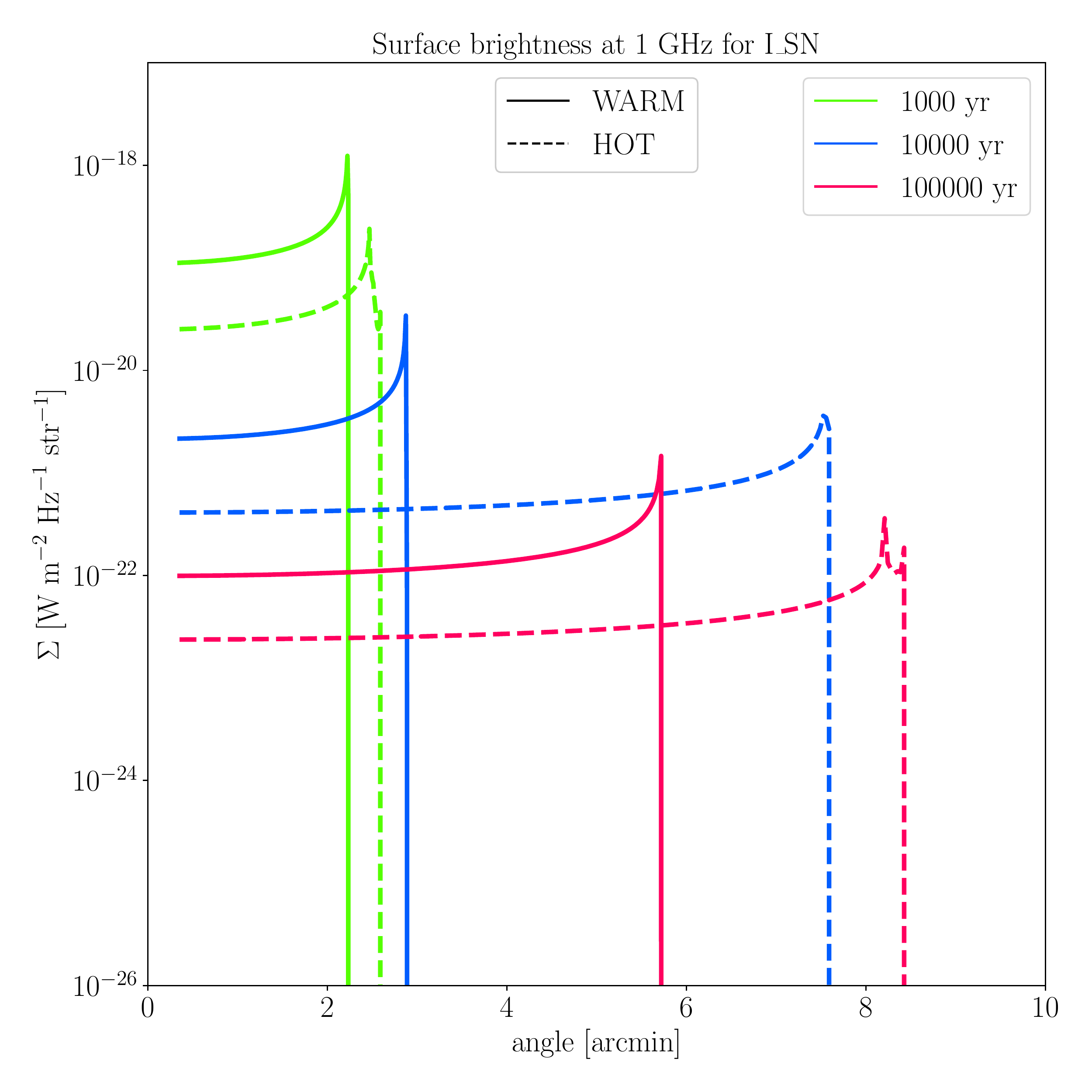}
\includegraphics[width=0.9\columnwidth]{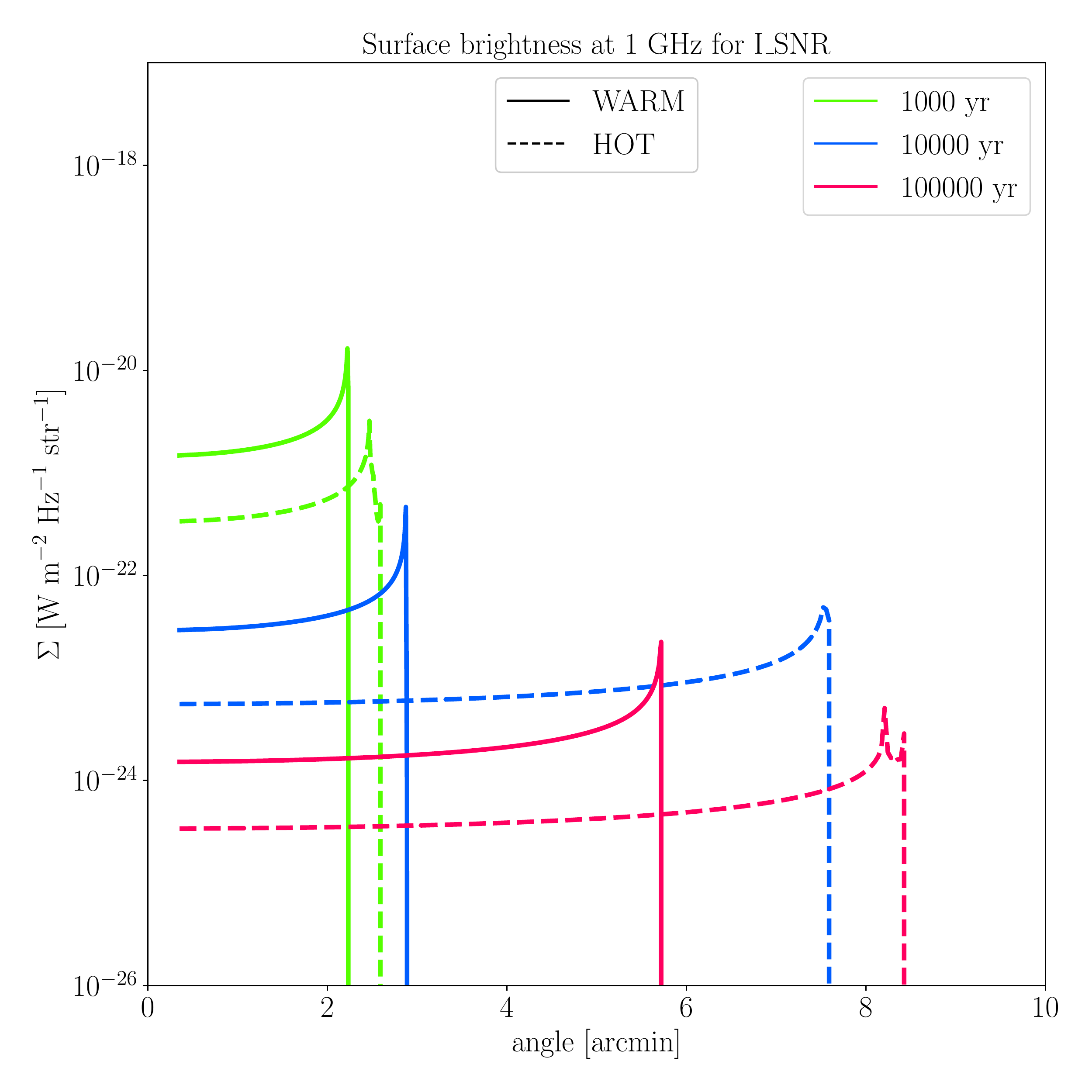}\\
\includegraphics[width=0.9\columnwidth]{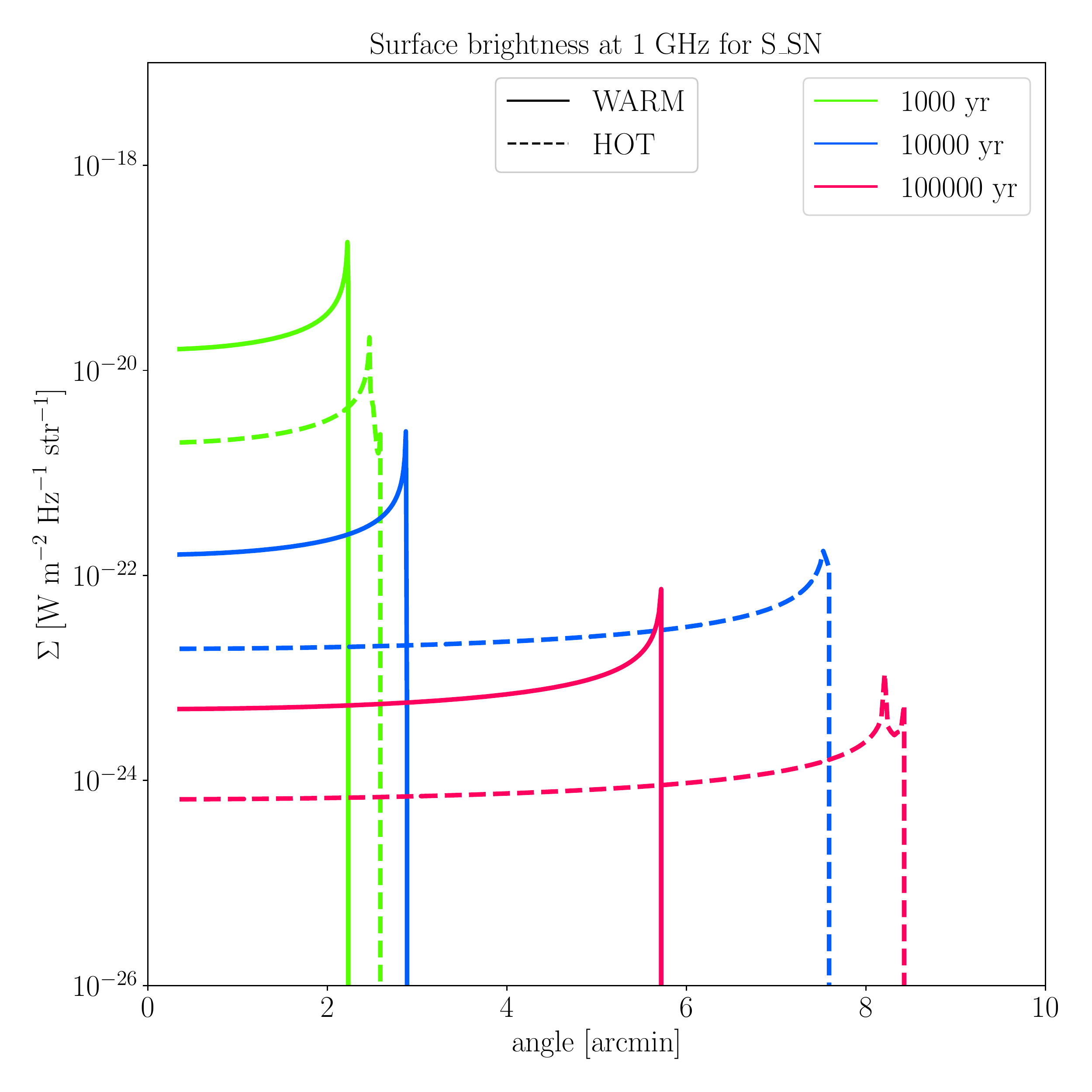}
\includegraphics[width=0.9\columnwidth]{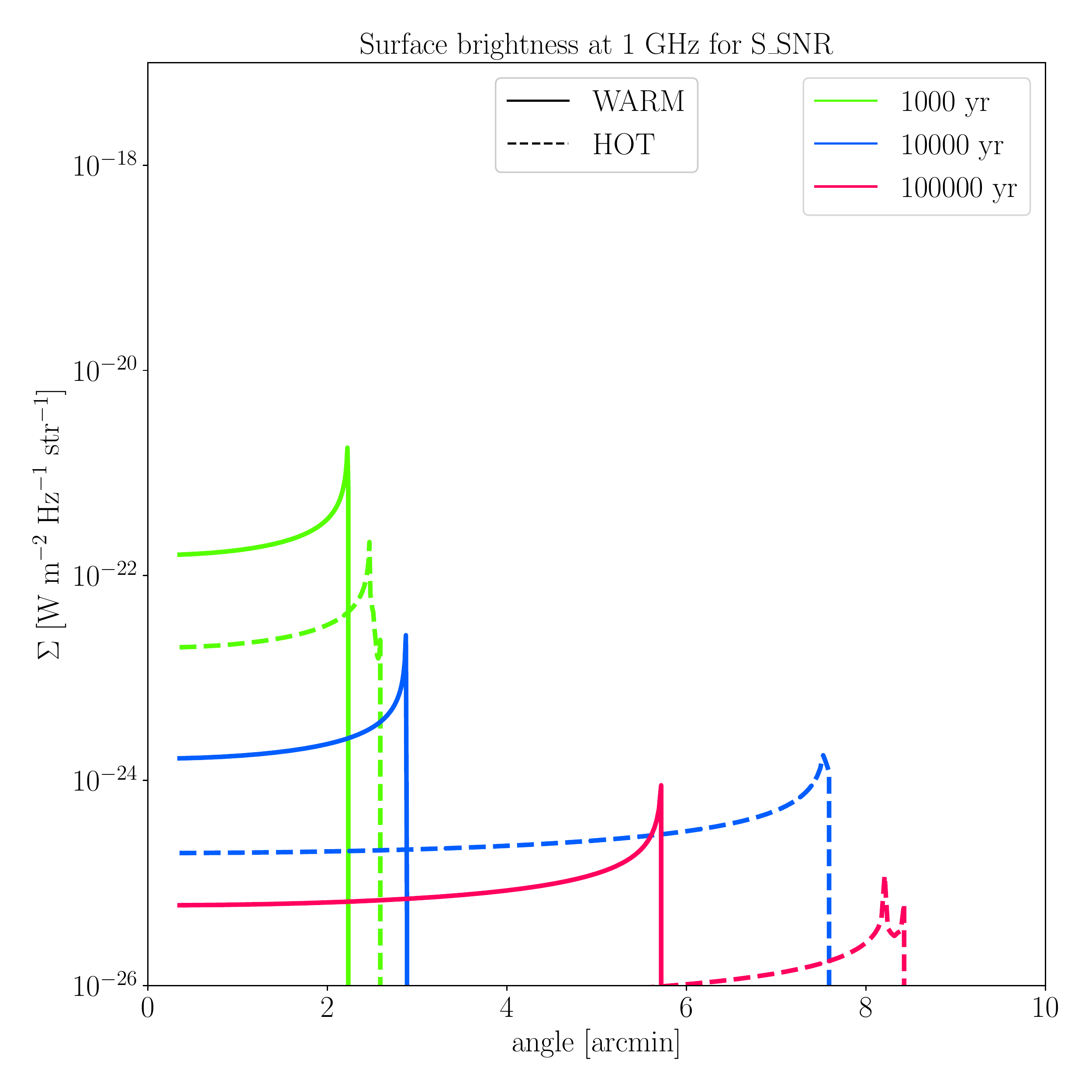}
\caption{Time evolution of surface brightness as a function of sky projection angle at 1 GHz. The colors of the curves depict the time evolution.}
\label{fig:Sigmatheta}
\end{figure*}

\begin{figure*}
\centering
\includegraphics[width=0.9\columnwidth]{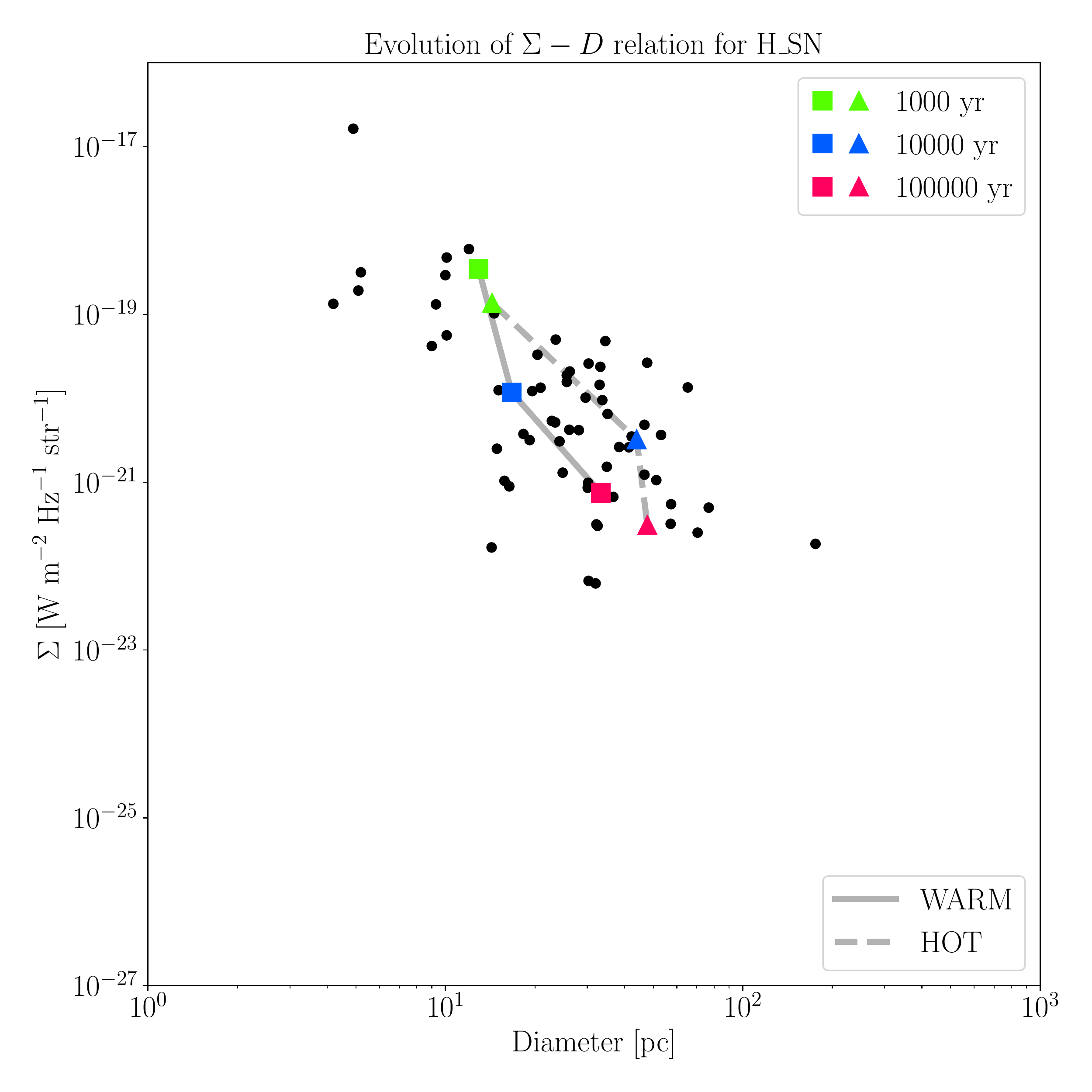}
\includegraphics[width=0.9\columnwidth]{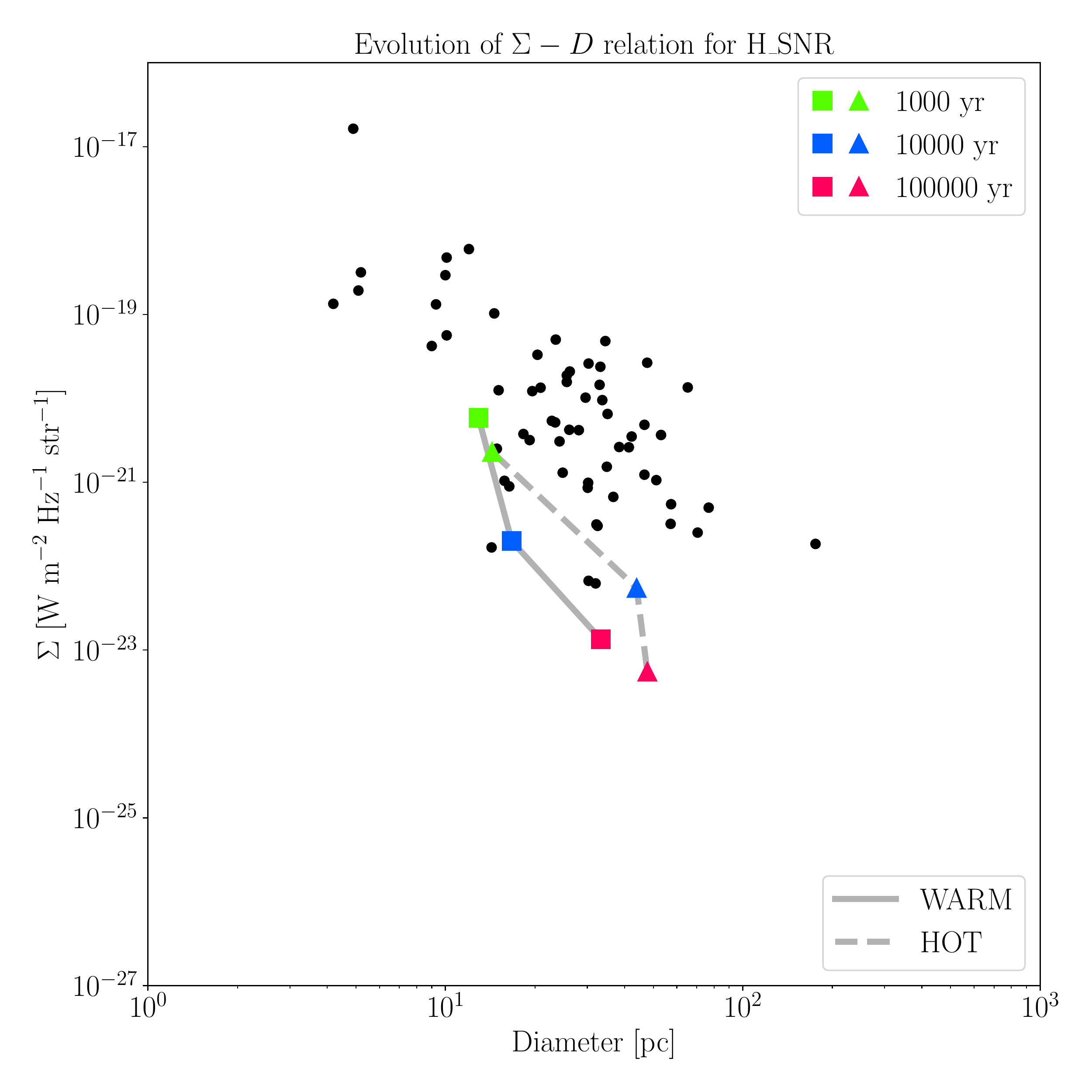}\\
\includegraphics[width=0.9\columnwidth]{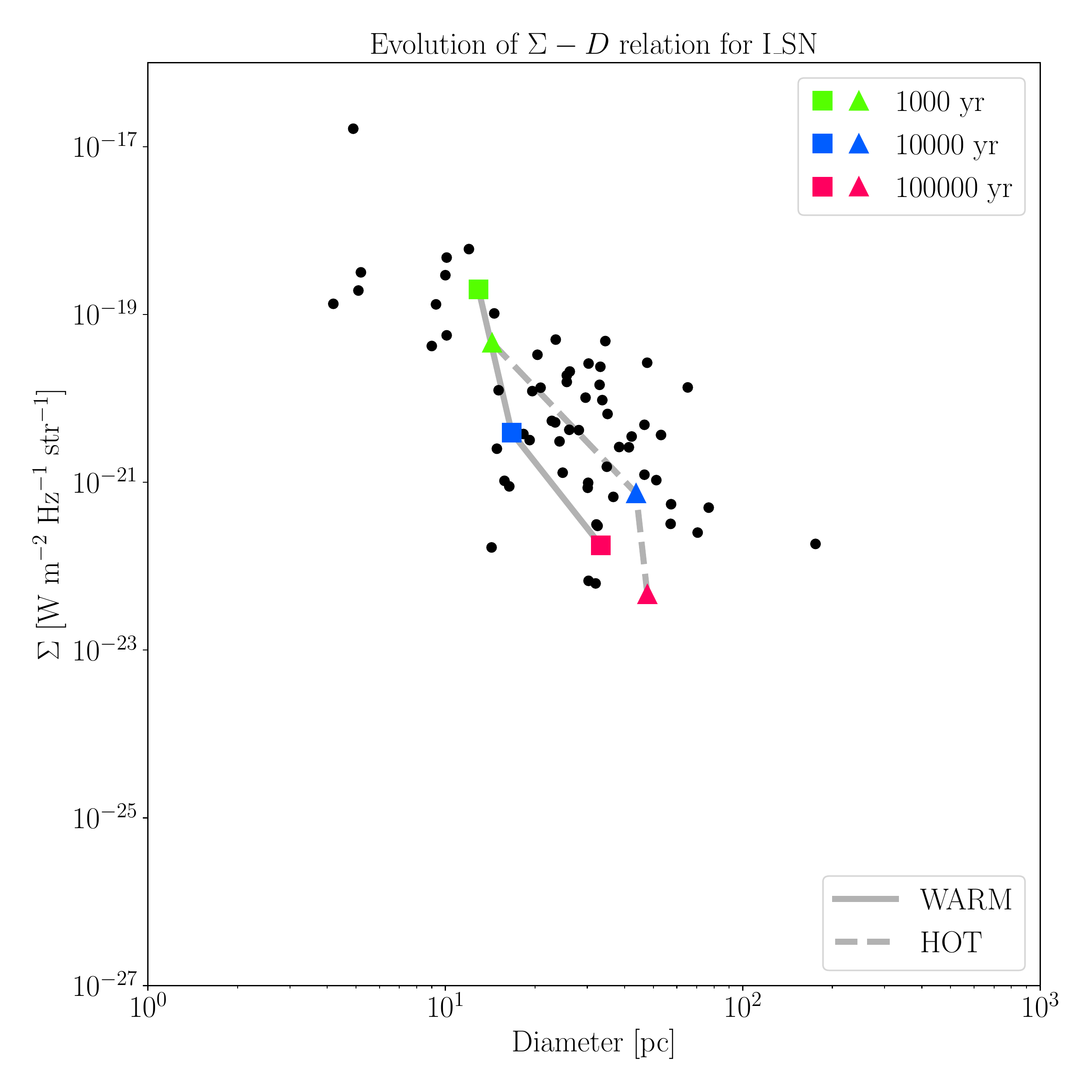}
\includegraphics[width=0.9\columnwidth]{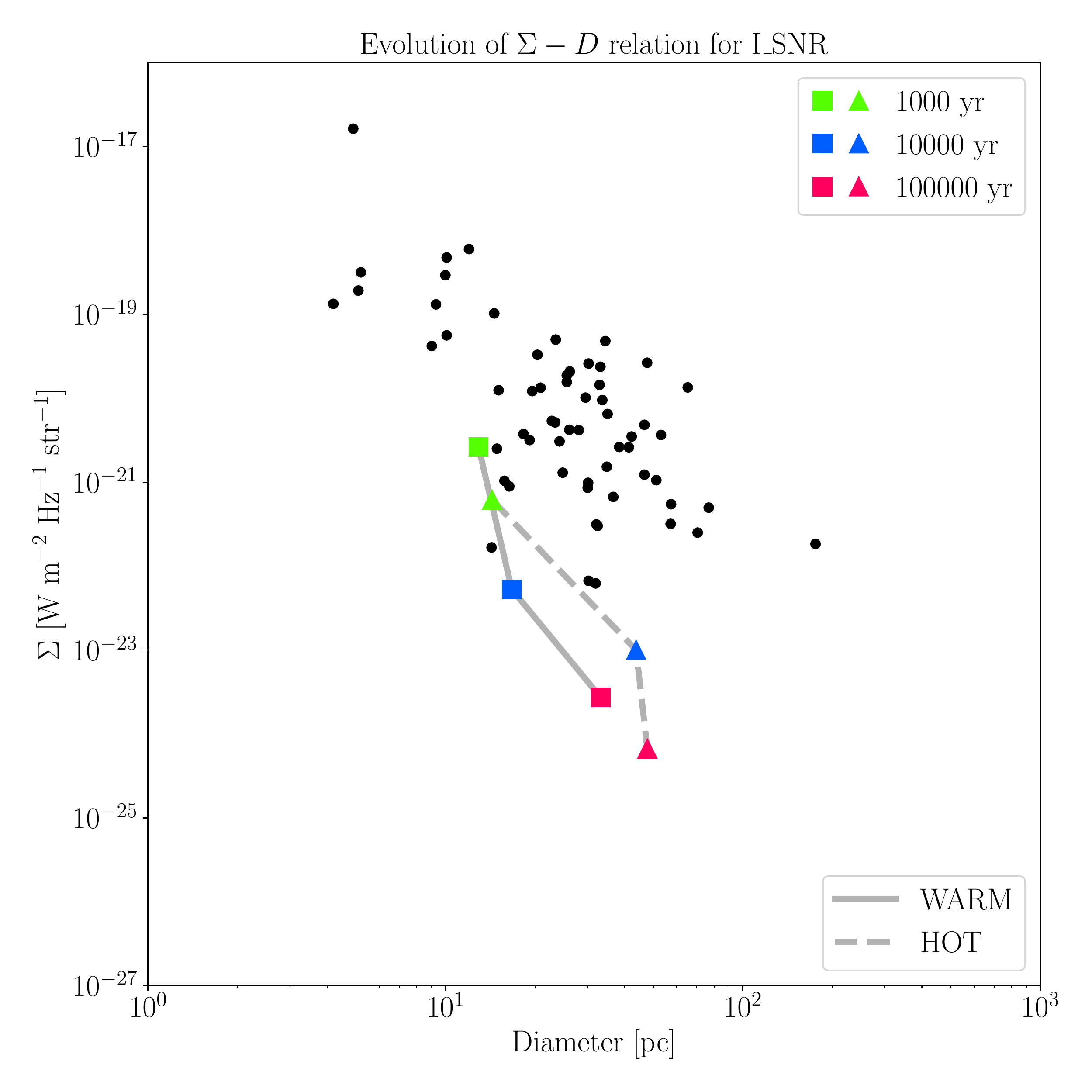}\\
\includegraphics[width=0.9\columnwidth]{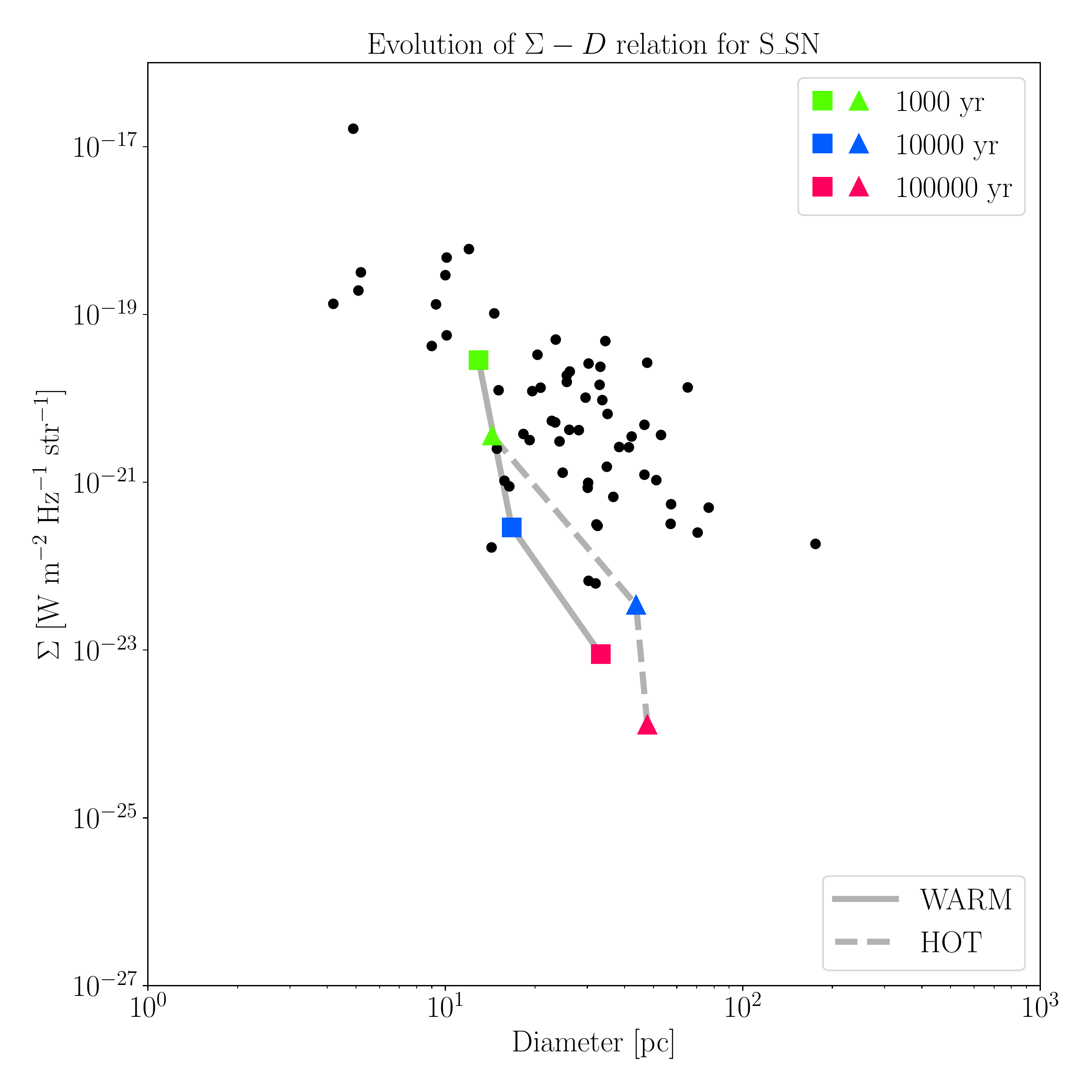}
\includegraphics[width=0.9\columnwidth]{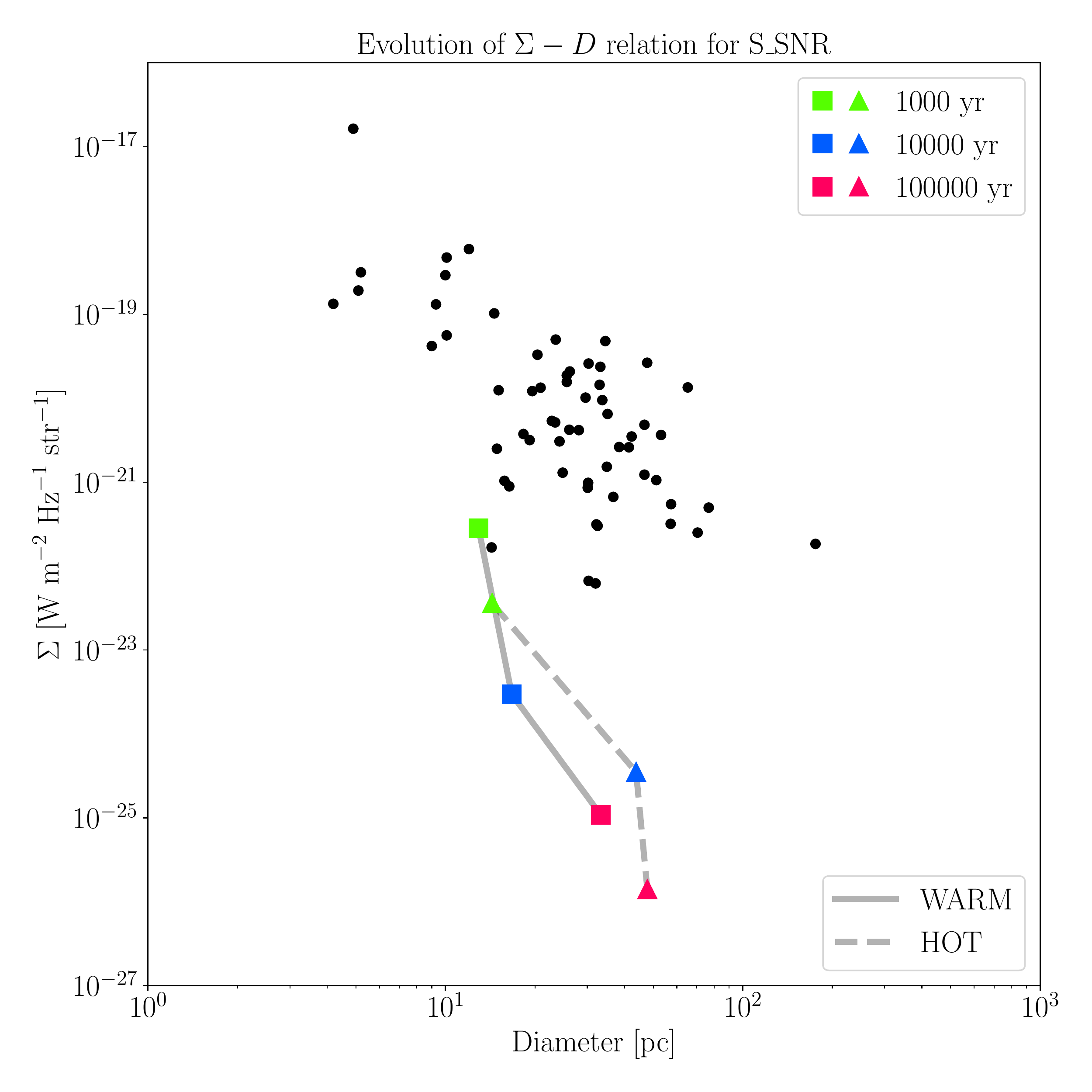}
\caption{Time evolution of the surface brightness from our models are plotted as a function of the SNR diameter. The colored data points connected by solid lines show the time evolution for our model. The black points represent a selection of observed Galactic SNRs as summarized in \cite{2013ApJS..204....4P}.}
\label{fig:SigmaD}
\end{figure*}

Firstly, we discuss the hydrodynamics of the interaction between the USSN ejecta and the CSM. In Figure~\ref{fig:rho_vel}, the time evolutions of the density and velocity profile are shown. {Here we mention on the dependence of the density profile on the ISM state. We can see that the model `HOT' has a larger radius of the ISM wall than the model `WARM', even though these two models have initially the same pressure. This suggests that the ISM density is important for dictating the location of the ISM wall; the lower ISM density (hot ISM) allows the exploding SNR gas to further expand. This feature is critical for quantifying the surface brightness of the USSNRs (see Figure~\ref{fig:Sigmatheta} and Figure~\ref{fig:SigmaD}).} 

From the density distributions, we can see that the ejecta keeps expanding until $t\sim 10000\,$years but starts decelerating around the ISM wall. The system can expand further {for another $\sim 3$ and $10\,$pc at most from the location of the ISM wall in the model `WARM' and `HOT', respectively}. {This can be observed in the panel of the velocity profile; the system experiences fast expansion at $t\lesssim 3000\,{\rm years}$, while after the collision with the ISM wall it only possesses several hundreds ${\rm km\ s}^{-1}$ of the outward velocity.} This implies that the diameter of the USSNR is highly constrained by the location of the ISM cavity wall, which in turn depends on the pre-SN mass loss activity of the progenitor. This picture can be applied to all core-collapse SNRs in general, for which the diameters of SNRs are associated with the pre-SN mass loss activity of their progenitors \citep[e.g.,][]{2021arXiv210904032Y, 2021arXiv211109534Y}.

\begin{figure}[ht!]
\centering
\includegraphics[width=0.9\columnwidth]{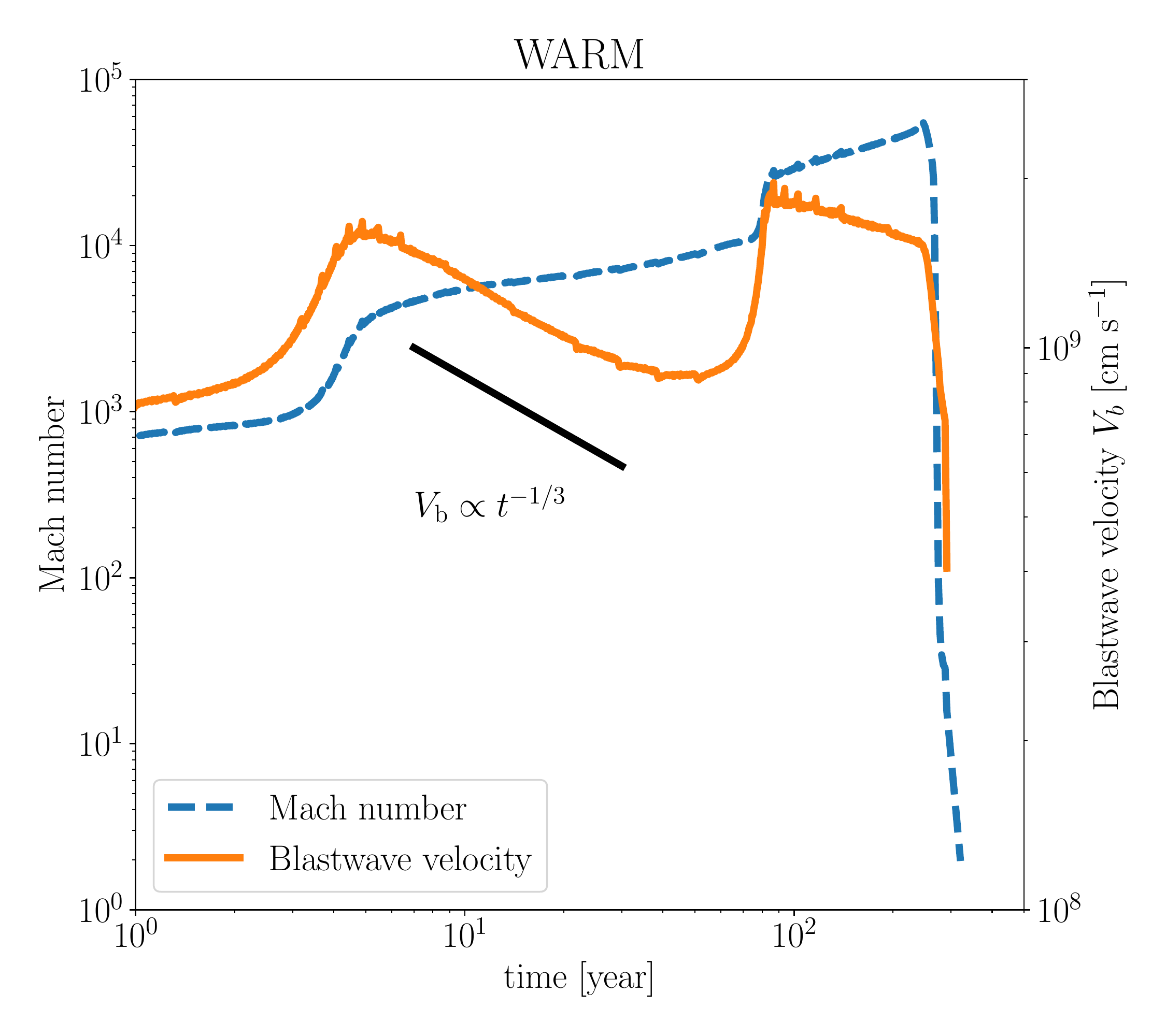}\\
\includegraphics[width=0.9\columnwidth]{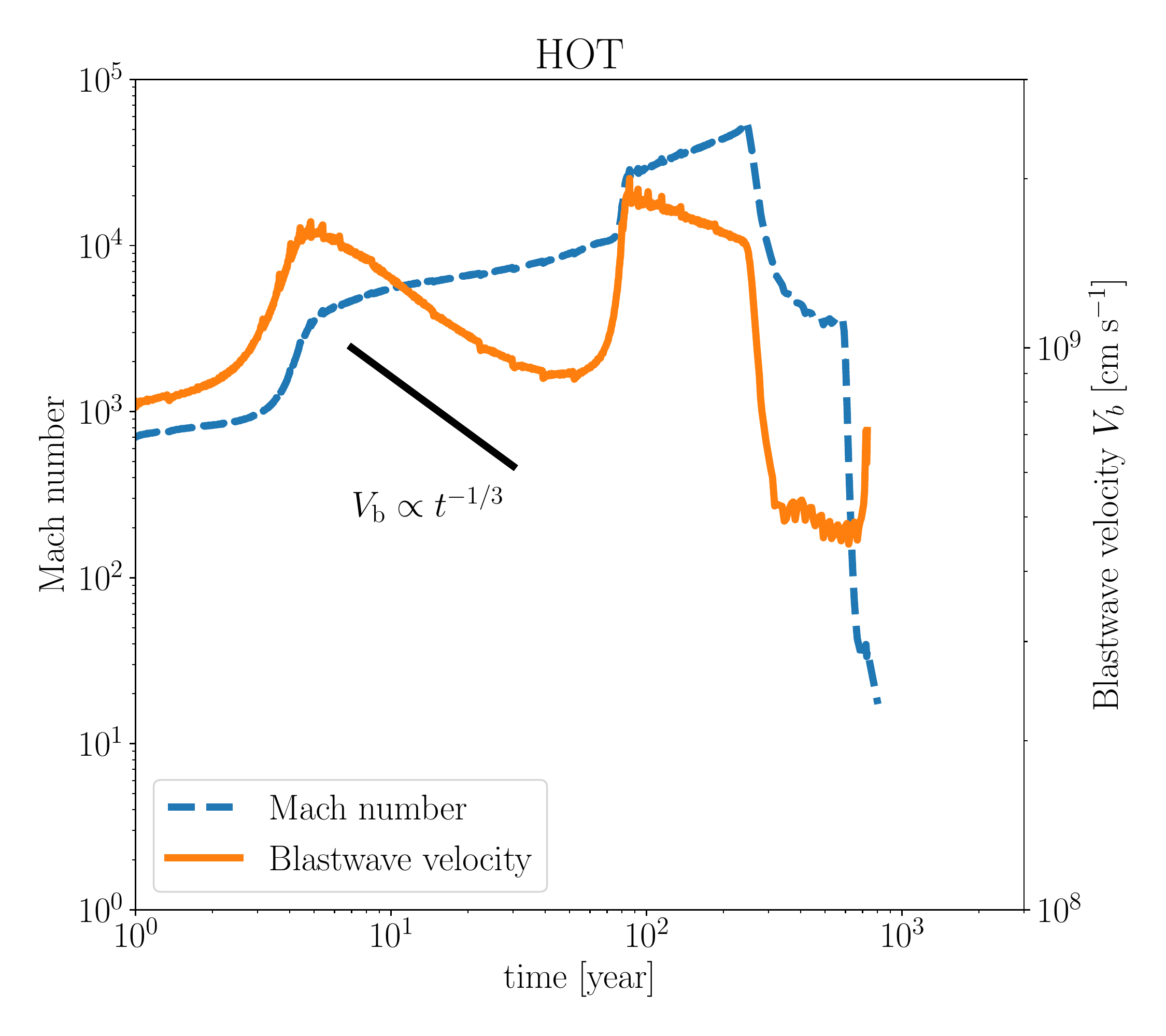}\\
\caption{Time evolution of the Mach number (dashed blue) and velocity (solid orange) of the blastwave in the model `WARM' (top) and `HOT' (bottom). After the steep drops of the blastwave velocity near the end of the curves, the shock Mach number drops to below 3 where we truncate the simulations. The black line shows the time dependence of the blastwave velocity expected in the Sedov phase, indicating a good agreement with the numerical solution.}
\label{fig:blastwave}
\end{figure}

Figure~\ref{fig:blastwave} shows the time evolution of the {Mach number and} the blastwave velocity. Within the first $300\,$years{, these two quantities both in the model `WARM' and `HOT' behave similarly each other since in this phase the identical CSM structure is traced. We} can see two epochs at which the blastwave accelerates at $r\sim0.01\,$pc and $r\sim 0.1\,$pc respectively, where the CSM density drops by orders of magnitude. {Correspondingly, the Mach number also increases by more than an order of magnitudes at the same time.} Overall, the velocity stays at about $10^9\,$cm\,s\pow{-1}, leaving the USSNR active for the first $300\,$years. Furthermore, at $5\,{\rm years} \lesssim t \lesssim 50\,$years when the swept CSM mass begins to exceed the ejecta mass, the velocity of the blastwave decays proportional roughly to $t^{-1/3}$. This agrees with the expected time dependence of the velocity in the Sedov phase for a CSM density profile proportional to $r^{-2}$ \citep{1994Bookshock}. {The gradual increase of the Mach number during that phase can be also observed, due to the decrease of the upstream temperature (see Figure~\ref{fig:CSMmodel_temp})}.

After $t\sim 300\,$ years, the blastwave decelerates down to $\sim 10^8\,{\rm cm\ s}^{-1}$, and then simply disappears out, as well as the Mach number decreases rapidly down to $\mathcal{O}(1)$. {This phenomenon can be observed both in `WARM' and in `HOT' though there are some quantitative differences between these two models.} This is caused by the hot plasma at $r \sim 5\,$pc shown in Figure~\ref{fig:CSMmodel_temp}; as the blastwave plunges into the plasma where the sound speed is high, the Mach number of the blastwave quickly decreases down to unity. It is implied that such a weak shock cannot support an efficient DSA. Additionally, the density jump at $r\sim 3\,{\rm pc}$ can also give rise to the deceleration of the blastwave. In conclusion, this result indicates that the blastwave in a USSNR dies out by propagating into a region of hot plasma at $\lesssim 10^3\,$years.

\begin{figure*}[ht!]
\centering
\includegraphics[width=0.9\columnwidth]{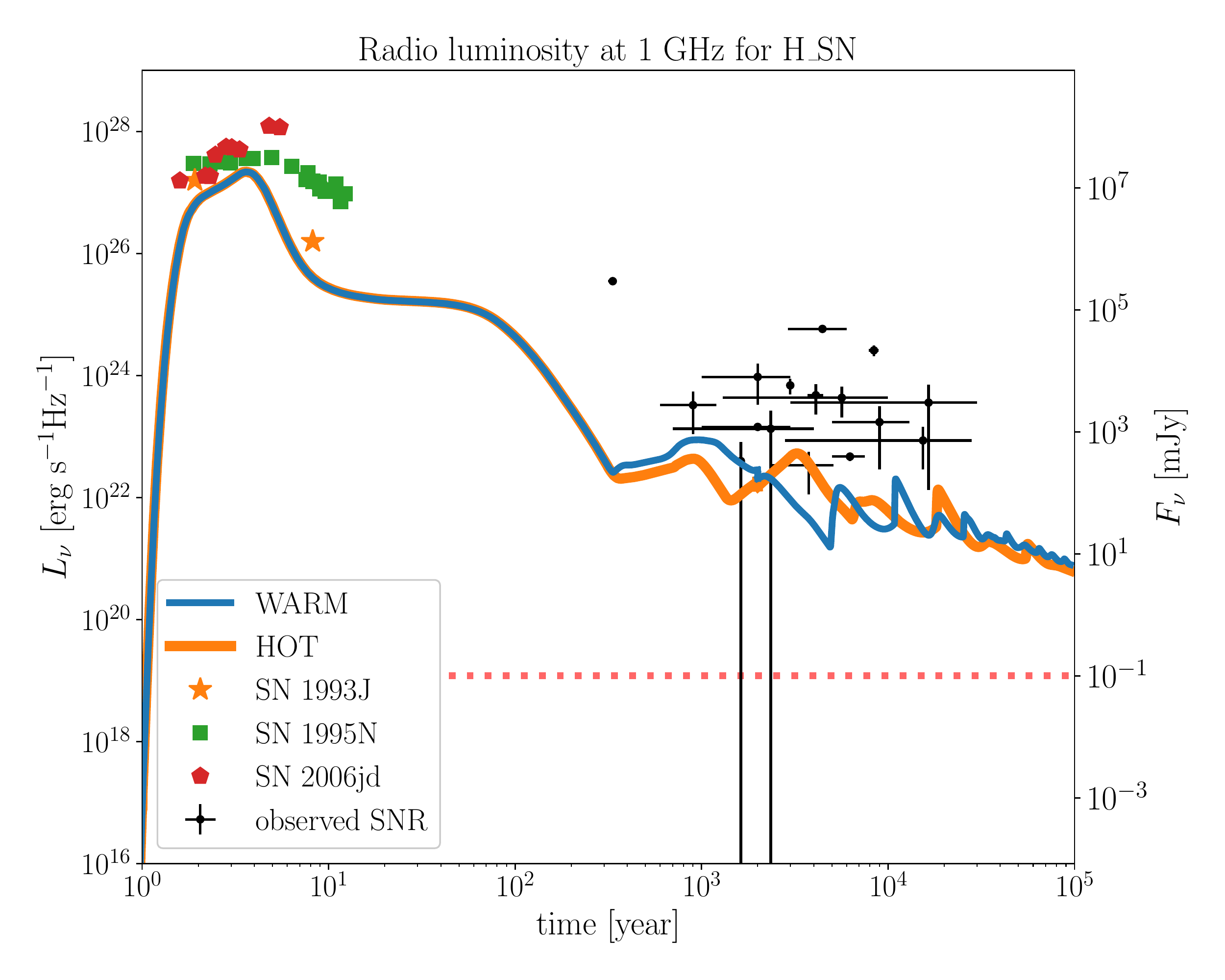}
\includegraphics[width=0.9\columnwidth]{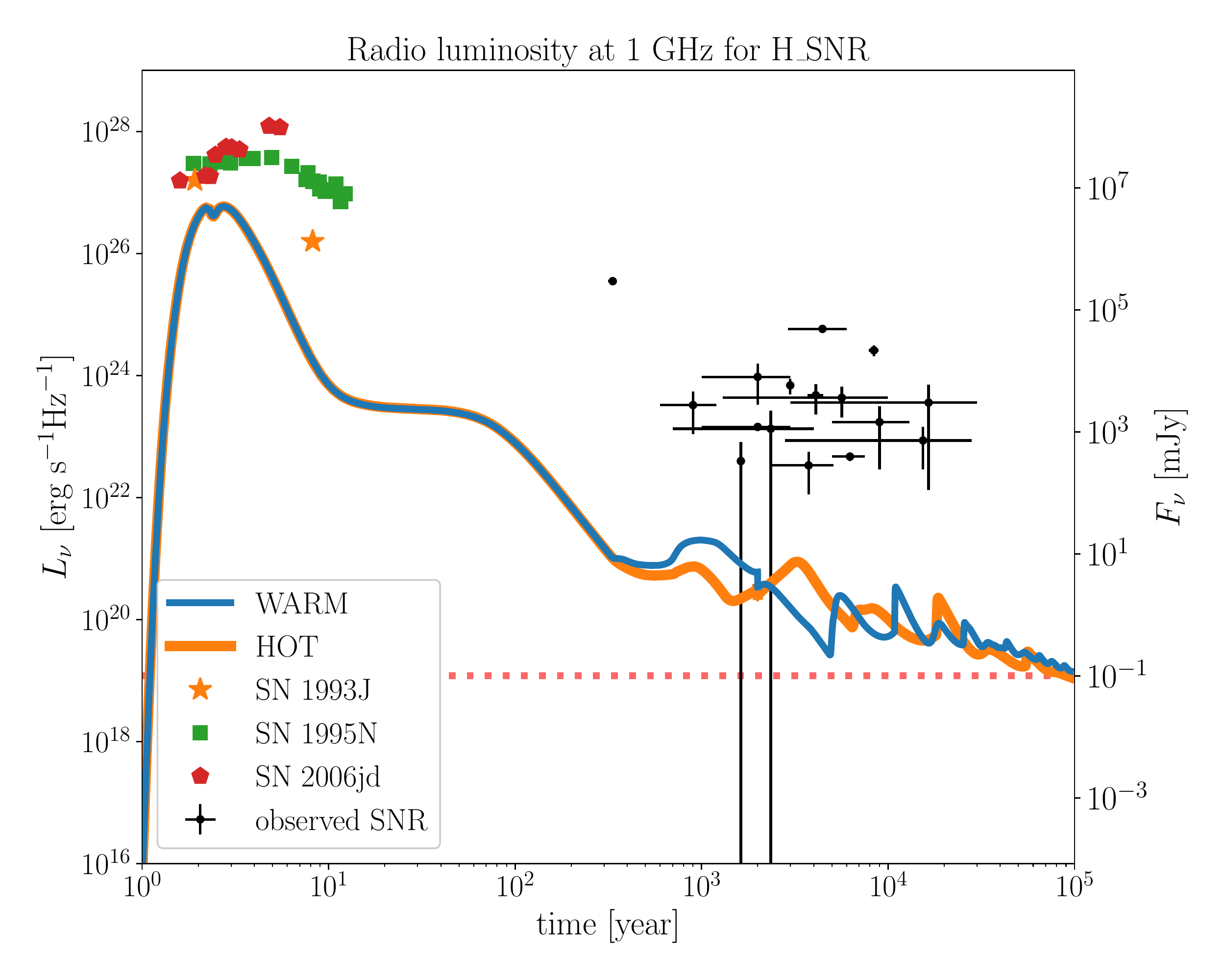} \\
\includegraphics[width=0.9\columnwidth]{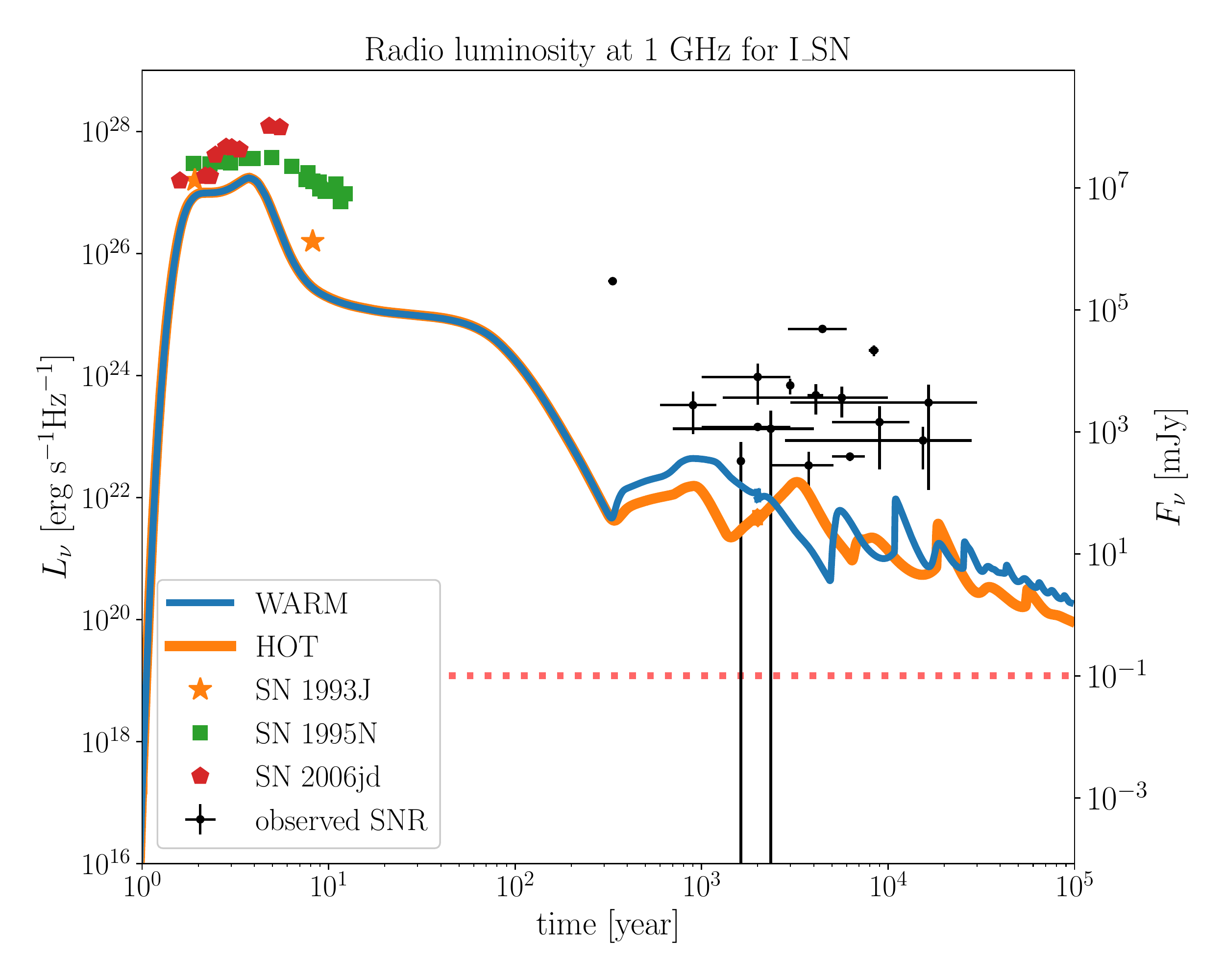}
\includegraphics[width=0.9\columnwidth]{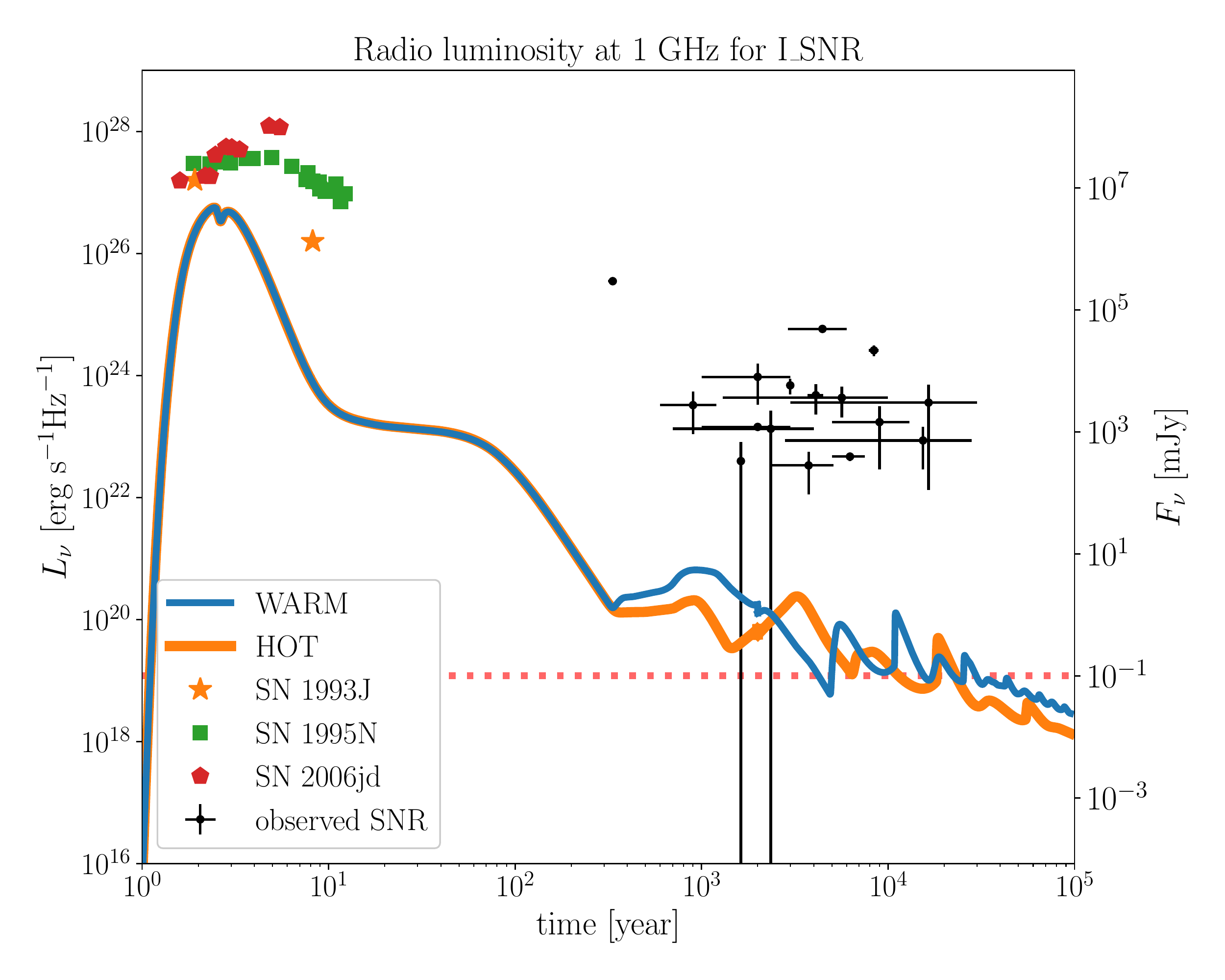} \\
\includegraphics[width=0.9\columnwidth]{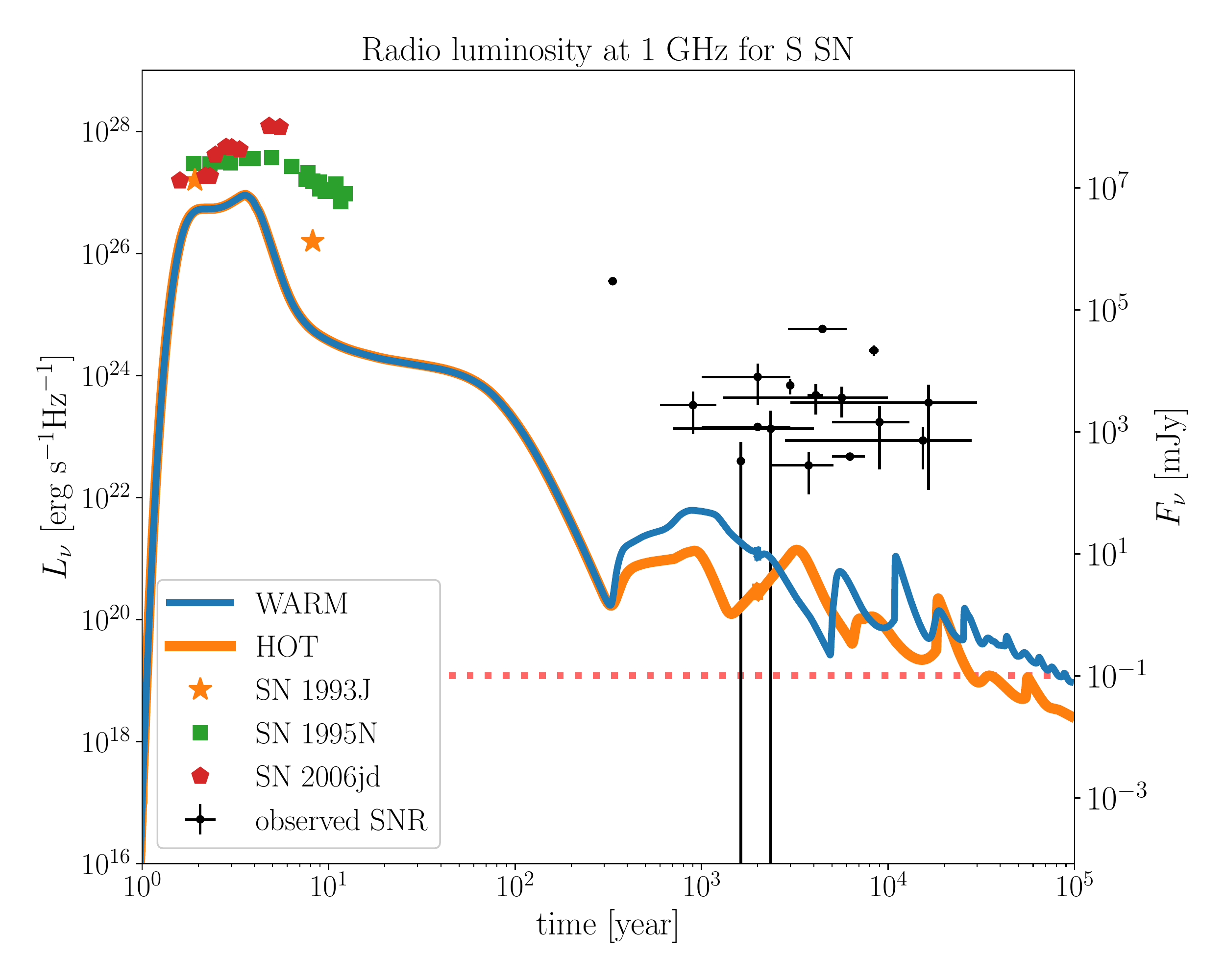}
\includegraphics[width=0.9\columnwidth]{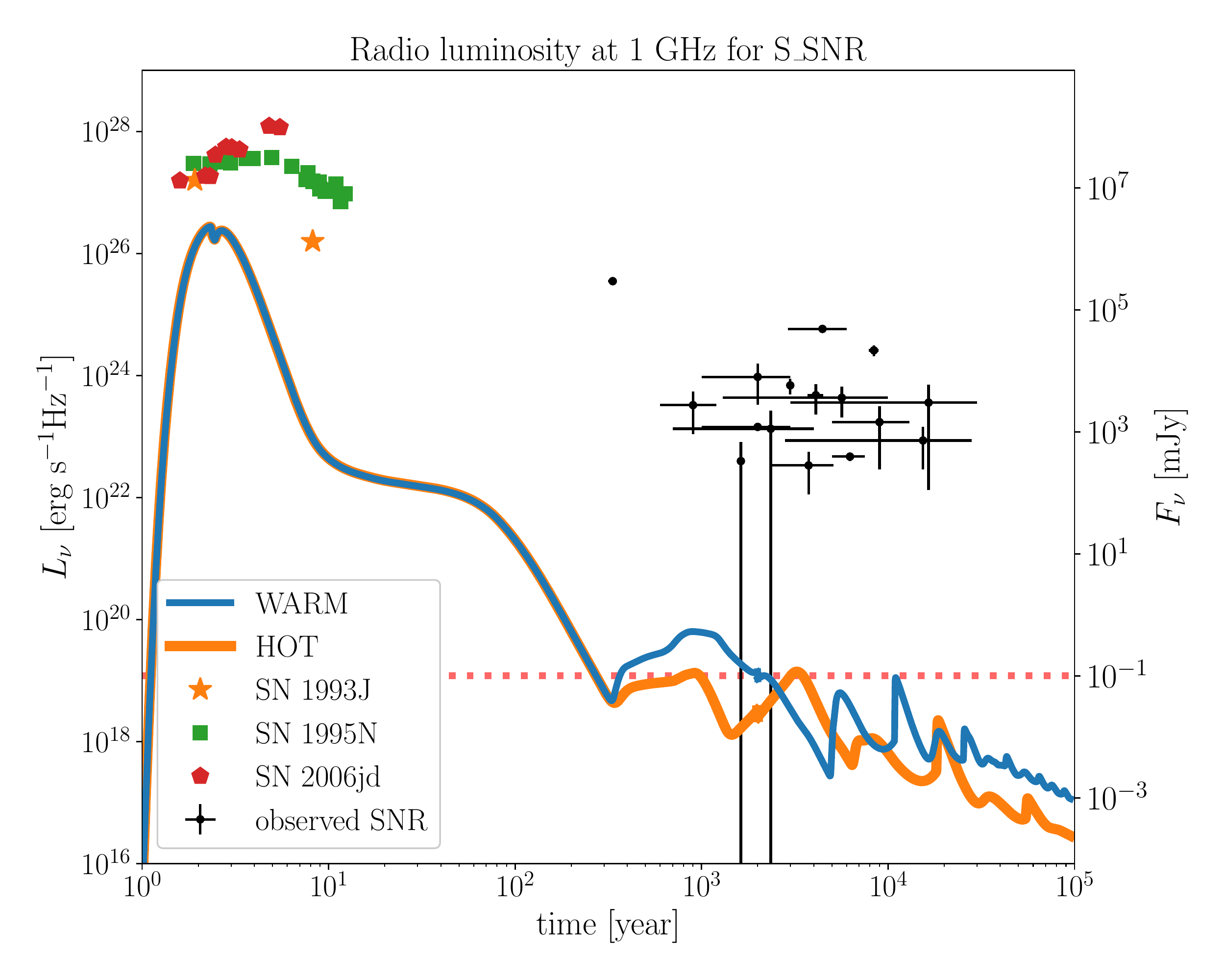}
\caption{Long-term radio light curves at 1 GHz compared to radio observations of SNRs. {Also plotted are the luminosities of SN 1993J (orange stars), SN 1995N (green squares), SN 2006jd (red pentagons), and Galactic SNRs listed in Table~\ref{tab:LCplots} (black points with error bars), estimated by the distances to each objects. The right y-axis stands for the observed flux densities with which the source with the luminosity shown in the left y-axis is observed at a distance $d=10\,$kpc.} The red dotted line indicates the detection limit of VLASS \citep[][]{2020PASP..132c5001L}.}
\label{fig:radioLC}
\end{figure*}

Figure~\ref{fig:radioLC} shows the long-term 1 GHz radio light curves from the models shown in Table~\ref{tab:parameters}. The observed flux density $F_\nu$ shown in the right y-axis is normalized by a distance $d=10\,$kpc. The peak luminosity of the light curve is determined by synchrotron self-absorption with their shapes slightly modified by free-free absorption \citep[see also][]{2019ApJ...885...41M}. Note that for {a USSN candidate iPTF14gqr non-detections of radio signals at the frequency $6\,$GHz and $22\,{\rm GHz}$ within $10\,{\rm days}$ have been reported, placing} upper limits \citep{2018Sci...362..201D}. In such a very early phase, free-free absorption completely damps the centimeter radio emissions, much more for $1\,{\rm GHz}$ \citep{2020ApJ...898..158M}. Since more electrons are accelerated and magnetic field is more intensively amplified in the models which assume larger efficiencies for DSA, brighter radio emission from USSNRs can be expected in the model with $(\epsilon_e, \epsilon_B) = (10^{-2}, 10^{-1})$ than those with $(\epsilon_e, \epsilon_B) = (10^{-3}, 10^{-2})$. Besides, a harder spectral index increases the number of more energetic electrons in the shocked region, which also results in the luminous radio signals. This behavior can be confirmed by comparing the luminosity between the models with $p=2.1$ and those with $p=2.5, 3.0$. {We note that there are no qualitative difference in the light curve behaviors between the two CSM models over the entire timespan up to $10^5\,$years. Actually as `HOT' has a more {extended} structure than `WARM' {as seen in Figure~\ref{fig:rho_vel}, a difference} between these two models is expected in their surface brightness as we will discuss later. 
} 

We first look at the behaviors of the young USSNR at ages less than $1000\,$years, {and compare} them with SNe well-observed at the frequency $\sim 1\,{\rm GHz}$ even 1 year after their explosions such as SN 1993J \citep{2011AA...526A.143M}, SN 1995N \citep{2009ApJ...690.1839C}, and SN 2006jd \citep{2012ApJ...755..110C}, and one of the youngest Galactic SNR Cas A \citep[][the point plotted at $t\sim 300\,$years]{2003ApJ...589..818D}. As seen in Figure~\ref{fig:radioLC}, our models show that young USSNRs at an age $t\sim 10\,$years and $t\sim 300\,$years produce fainter radio signal than those from the bright SNe and Cas A, respectively. The relatively weak emissions can be partially attributed to the shock velocity which is by a factor of a few lower than what is inferred for these objects \citep[see, e.g.,][]{1998ApJ...509..861F}. Another possible reason is that at $t\sim 100\,$years the blastwave is propagating at $r\sim 1\,$pc where the dense CSM formed by the mass loss driven by the RLO is absent. Then the density of the CSM swept by the blastwave is considerably small there, making the DSA less efficient. However, we note that the {expected} flux density of the radio emission from the USSNR at $d = 10\,$pc {keeps} greater than $0.1\,$mJy within an age $t\lesssim 1000\,$years, which is bright enough to be detected by the present radio surveys such as Very Large Array Sky Survey \citep[VLASS,][]{2020PASP..132c5001L}, if it inhabits inside our galaxy.

\begin{figure}[ht!]
\plotone{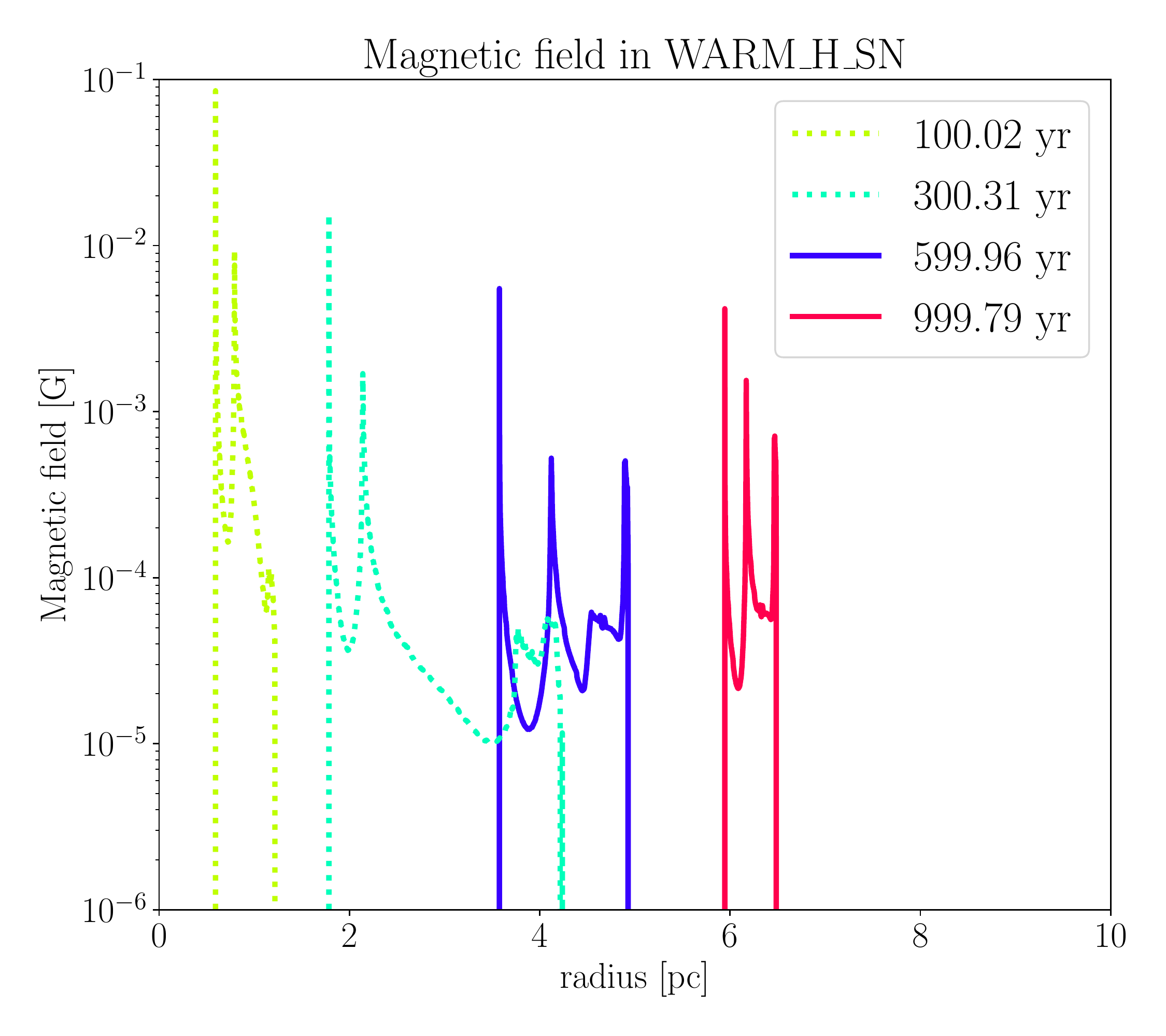}
\caption{{Time evolution of the profile of the magnetic field in the model `WARM\_H\_SN'. The development of the strength of the magnetic field at the forward shock can be observed at $t\gtrsim 500\,{\rm years}$ (solid), rather than before (dotted).}}
\label{fig:jiba}
\end{figure}

Next we discuss the properties of the light curves of USSNR at larger ages ($1000\,\mbox{years} \lesssim t \lesssim 10^5\,\mbox{years}$). At $t\sim 1000\,$years, the radio emission brightens by a factor to an order of magnitude compared to $t\sim 300\,$ years, even though the synchrotron emission in this phase is optically thin to self-absorption. This enhancement stems from the interaction between the SN ejecta and the relatively dense CSM located at $\sim 3$--$10\,$pc; {a larger amount of the gas injection into the shocked region leads to a larger number of the synchrotron {emitting} electrons, resulting in a higher radio luminosity. In addition, the compression of the gas around the blastwave by the collision with the dense CSM brings about the further amplification of the magnetic field through the conservation of the magnetic flux {(see Figure~\ref{fig:jiba})}. This can also be a cause of the brightening of the radio luminosity.} We note that this {brightening} is one of the characteristics of a USSNR associated with the time dependent mass loss driven by RLO, since a CSM with a simple power-law distribution cannot reproduce such a rise in radio luminosity in the optically-thin regime. Yet, the subsequent radio signals are fainter than those observed from the Galactic SNRs enumerated in Table~\ref{tab:LCplots}. The stalled blastwave at $t\sim 300\,$years can no longer execute efficient DSA any further. Even so, it is worth mentioning that SNRs discovered so far are biased towards \deleted{the }bright objects. Deep surveys such as VLASS will have potential to uncover the population of the SNRs as faint as the aged USSNRs.

After the death of the blastwave, DSA will no longer be triggered, and the non-thermal emissions are forced to decline through adiabatic cooling. The timing of dominance by adiabatic cooling is roughly $1000\,$years, and is more-or-less determined by the location of the hot plasma (Figure~\ref{fig:CSMmodel_temp}). The hot plasma is formed by the interaction between the He-rich wind blown from the progenitor binary and the H-rich gas originated from the CE ejection or the uniform ISM. Our result implies that the location of the hot plasma in the CSM is key to determining the lifetime of the blastwave and hence the observable lifespan of the USSNR.

We also observe oscillations of the light curves at $t\gtrsim 10^4\,$years. This is an one-dimensional artifact due to the reflective condition at the inner boundary of the simulation domain. As the reverse shock of the USSNR brings along an inward gas flow back to the explosion center, it rebounds back to the outer interacting region. Then the material around the shocked region is compressed, inducing an amplification of the magnetic field through flux conservation. A repeating occurrence of this inward and outward motion results in the oscillation of the radio luminosity in our models for the aged USSNR. In practice, multi-dimensional dynamics should suppress the motion of the gas described above due to a broken spherical symmetry. Even so, it can be noted that the global evolution of the radio luminosity of the aged USSNR roughly follows an adiabatic evolution when averaged over a longer timescale.


Figure~\ref{fig:Sigmatheta} shows the time evolution of the surface brightness as a function of the sky projection angle. {The model `HOT' has fainter surface brightness and larger projection angles at which the surface brightness becomes maximum ($\theta_{\rm max}$) than those in the model `WARM', because the model `HOT' has a more extended CSM density structure than the model `WARM' (see Figure~\ref{fig:CSMmodel_rho}). Yet the qualitative behavior of the surface brightness as a function of the sky projection angle is similar between these two models. }$\theta_{\rm max}$ is mainly dictated by the location of the ISM wall, which prevents the gas in the shocked region from expanding any further outward (see Figure~\ref{fig:rho_vel}). As mentioned before, the hot plasma and the ISM cavity wall are shaped by the wind colliding with the CE {and/}or the \deleted{uniform }ISM, which ultimately determines the detectability of the USSNR.

The evolution of the relation between the surface brightness and diameter of the USSNR can be assessed by the $\Sigma$-$D$ diagram shown in Figure~\ref{fig:SigmaD}. {For the same reason as the relation between $\Sigma$ and $\theta_{\rm max}$ (Figure~\ref{fig:Sigmatheta}), the model `HOT' has a fainter surface brightness and larger diameter than the model `WARM'. This results in the lower right position of the evolutionary path of the model `HOT' in the $\Sigma-D$ diagram.} The magnitude of the surface brightness strongly depends on the parameters relevant to the DSA (i.e., $p, \epsilon_e$, and $\epsilon_B$). The surface brightness of the model appears to be relatively faint compared to those of the Galactic SNRs {in the models such as `H\_SNR', `I\_SNR', `S\_SN', and `S\_SNR'}, in which the expected flux density of the radio emission from the aged USSNRs are approximately $0.1$ mJy{. This poses a challenge to detection and is} consistent with the current non-detection of the SNR hosting a DNS binary in our Galaxy. On the other hand, {in all of our models} the USSNR diameter is in the order of $10\,$pc, which is also typical of the observed Galactic SNRs \citep{2013ApJS..204....4P}. We suggest that a faint surface brightness combined with a diameter $D \sim 10\,$pc can be a characteristics of a USSNR, and might be useful diagnostics for searching SNRs hosting a DNS binary.

{At last we comment on the role of the ISM state on the radiative characteristics of the\deleted{following} USSNRs. Comparisons of the solid and dashed lines\deleted{ with the same colors} in Figure\,\ref{fig:Sigmatheta} and Figure\,\ref{fig:SigmaD} demonstrate that the surface brightness of `HOT' is fainter than that of `WARM' at the same age. This is attributed to the fact that `HOT' has a larger diameter and sky-projected angular size than `WARM' and that the luminosities of these two models are similar to each other. Our simulations of CSM formation {(Section\,\ref{sec:CSMformation}) assume that the models `WARM' and `HOT' have} the same thermal pressure but different densities in the initial profiles; we have shown that the model with a lower initial density leads to a larger diameter of the USSNR. Thus we conclude that the ISM density plays a role in determining the physical scale of the USSNR, which also affects the surface brightness.}

\section{discussion}\label{sec:discussion}
\subsection{The USSNR population}\label{sec:population}

Equipped with our models, it is possible for us to make predictions for \deleted{the}general properties of the USSNR population. Two timescales are important for characterizing the SNR population. {One} is the observable lifespan of the SNR, $t_{\rm snr}$, defined here as the timescale in which the radio emission from the SNR can be detected. The other one is $t_{\rm sn}$, the time interval between subsequent SNe or the inverse of the SN rate in a galaxy. The number of active SNRs can then be estimated as $t_{\rm snr}/t_{\rm sn}$. As for USSNRs, \cite{2019ApJ...882...93H} predicted the event rate of USSNe as $510.88\,{\rm gal}^{-1}\,{\rm Myr}^{-1}$ in their feasible population synthesis model, leading to $t_{\rm sn} \sim {2}\times 10^3\,$years\footnote{\cite{2019ApJ...882...93H} defined a USSN as an explosion of the star with its helium envelop mass less than $0.2\,M_\odot$, and an iPTF\,14gqr-like USSN as a USSN with its ejecta mass $0.15-0.30\,M_\odot$ containing the helium component $0.003-0.01,M_\odot$. The event rate of USSNe is estimated to be $\sim 10$ times larger than that of iPTF~14gqr like USSNe, and we adopt the former value.}
\deleted{Other studies of population synthesis calculations also suggest the consistent outcomes with the above, some of which are still upside values \citep{2018MNRAS.474.2937C}}. For the SNR lifetime, $t_{\rm snr} \sim 100$--$ 10^5\,$years can be implied from our models depending on the DSA efficiencies and the spectral index of accelerated electrons. Hence, the expected number of active USSNRs can be derived as {$0.002-20$}.

These estimations involve uncertainties {from observational conditions (e.g., sensitivity)} as well as the DSA parameters. Models with high DSA efficiencies or hard power-law index for the accelerated electrons (e.g., `H\_SN' and `H\_SNR') {probably over-estimate the observable lifespan; }typical shock acceleration efficiency constrained by SNR observations are usually found to be lower than those inferred from the observations of radio SNe \citep{2012ApJ...750..156L}. Moreover, it has been suggested that the spectral index of the accelerated particles in young SNRs can be modified and steepened by non-linear effects associated with magnetic field amplification in an efficient DSA and Alfv\'{e}nic drift effect \citep{2006ApJ...648L..33V, 2008AIPC.1085..336Z, 2019ApJ...876...27Y}, whereas in mature SNRs it tends to follow the prediction by the standard DSA \citep{2015MNRAS.451.3044R}. The former is more appropriate for the situation considered in the present work, since our simulations indicate that the blastwave dies out at a young age in our CSM model. From these arguments, we can refer `I\_SNR' as our fiducial models for the evolution of a USSNR, which predicts an observable lifespan $t_{\rm snr} \sim 10^4\,$years. Then we can further constrain the expected number of the observable USSNRs to be {$\sim 2$}. Since the detected number of the Galactic SNRs reaches $\sim 400$ \citep{2019JApA...40...36G}, the most probable fraction of USSNRs is then {at most} {$\sim 0.5$~\%} of all active SNRs. We note however that the quantification of the observable lifespan of the USSNRs involves uncertainties and depends on the sensitivity of the detectors as well.

{The expected number of active USSNRs in a galaxy, $\sim 2$,} poses a severe challenge on the search of USSNRs. Radio observation facilities capable of deep surveys such as the Square Kilometre Arrays (SKA) are requisite to solving this difficulty. A Galactic SNR survey with a sensitivity $\sim 0.1$\,mJy is {one of the solutions to search for USSNRs, as well as for eliminating the possible bias against faint SNRs. Another possible solution is to extend the search to other galaxies in the local group. SNRs producing radio emission brighter than $\sim 10^{21} (d/\mbox{Mpc})^2 (F_{\rm lim}/\mu\mbox{Jy})$ erg s\pow{-1} Hz\pow{-1} in the local group can be detected by making use of the deepest observation projects, where $F_{\rm lim}$ is the maximum sensitivity of SKA \citep{2019arXiv191212699B}. This sensitivity enables us to detect the radio emission from USSNRs (proved by the model `H\_SN' and `H\_SNR' in Figure~\ref{fig:radioLC}). Assuming that the galaxies in the local group have the same proportion of USSNRs to all kinds of SNRs ($\sim 0.5$\%), this attempt might offer an opportunity to discover USSNRs.}

\subsection{{General implications for} stripped-envelope SNRs}\label{sec:SESNRs}
We have shown that the blastwave of a USSNR suddenly loses its punch by being blunted by the hot plasma. The lifetime of the blastwave is limited to $\lesssim 10^3\,$years, and the diameter is roughly a few $10\,$pc. The evolution of USSNRs is different from that elucidated classically. Generally, after the Sedov phase at $\sim 10^4\,$years, radiative cooling from the swept ISM drains the internal energy away from the system, leading to a fast deceleration of the blastwave. Through the pressure-driven snowplow phase and momentum-driven snowplow phase, the SNR merge with the surrounding ISM at $t\sim 5\times 10^5\,$years \citep{1988ApJ...334..252C}. On the other hand, the evolution of USSNRs is heavily influenced by the non-uniform CSM density distribution and the presence of a hot plasma in the vicinity of the ISM wall, both of which are attributed to the wind driven by the binary interaction. {The binary interaction is a key physical process in the evolutionary behaviors of USSNRs that deviate from the classical picture of SNR evolution.}

Besides USSNe, it is widely believed that stripped-envelope SNe (Type IIb, Ib, and Ic SNe) are explosions of a massive star involved in binary interaction \citep[e.g.,][]{2010ApJ...725..940Y, 2017ApJ...840...90O, 2019NatAs...3..434F}. It can be speculated that the evolution of SNRs originated from stripped-envelope SNe also deviates from the classical theory. Considering that some fraction of the observed SNe are classified as stripped-envelope SNe \citep[Type IIb, Ib, and Ic SNe,][]{2013MNRAS.436..774E}, it is natural that some of the confirmed SNRs in our Galaxy also come from a stripped-envelope SN origin. {Previously, in terms of hydrodynamics, the effect of the wind bubble and its multi-dimensional behaviors on the subsequent SNR evolutions have been investigated by making use of simple models for stellar mass loading \citep{1990MNRAS.244..563T, 1991MNRAS.251..318T, 2005ApJ...630..892D, 2007ApJ...667..226D}, but models for mass loss history based on detailed binary evolution calculations have not been incorporated.} We thus suggest that such stripped-envelope SNRs should be modeled with the mass loss history of the progenitor binary taken into account for their surrounding CSM environments \citep[e.g.,][]{2021arXiv210904032Y, 2021arXiv211109534Y}.

{
\subsection{Radio emission from the hot plasma region}
\label{sec:radiofromwind}
The velocity of the RLO wind is {high, reaching $\sim 1000\,$km s\pow{-1}. It is therefore }possible that in the formation process of the hot plasma driven by the RLO wind, electron acceleration and magnetic field amplification can happen through the DSA mechanism. Such effects can contribute to the radio luminosity and surface brightness of the {subsequent USSNRs,} and thus an evaluation of this process is required. Our simulations show that the velocity of the RLO wind shock is $V_{\rm sh,RLO}\sim200$ km s\pow{-1}. If we consider a hot ISM state ($T_{\rm ism}\sim 10^6\,$K), the Mach number of the shockwave {launched by} the RLO wind is in the order of unity. This indicates that {the contribution from the hot plasma to the total flux of the radio emission from USSNRs is negligible in a hot ISM}. On the other hand, in a warm ISM ($T_{\rm ism}\sim 10^4\,$K), the Mach number is large enough to sustain DSA. The hot plasma can then be a potential emitter of synchrotron radiation. Assuming that the region is optically thin for synchrotron radiation, the radio luminosity is written as $L_\nu \sim 4\pi^2 R^3 j_{\nu, {\rm syn}}$, where $R$ is the position of the RLO wind shock. Based on the formulae introduced in this study, the luminosity can be roughly estimated as $L_\nu \sim 10^{21} R_{\rm 20pc}^3 \epsilon_{e,-3}\epsilon_{B,-2}^{3/4}\rho_{\rm ism,-24}^{11/8}V_{\rm sh,RLO,200}^{11/4}\,{\rm erg}\,{\rm s}^{-1}\,{\rm Hz}^{-1}$, where $R_{\rm 20pc}=R/({\rm 20pc}), \epsilon_{e,-3}=\epsilon_e/10^{-3}, \epsilon_{B,-2}=\epsilon_B/10^{-2}, \rho_{\rm ism,-24} = \rho_{\rm ism}/(10^{-24}\,{\rm g}\,{\rm cm}^{-3})$, and $V_{\rm sh,RLO,200}=V_{\rm sh,RLO}/(200{\rm km}\,{\rm s}^{-1})$. {Comparing this magnitude to the models, we can see that this contribution from the hot plasma in a warm ISM is negligibly small with respect to the predicted luminosities of young USSNRs ($t\lesssim 1000\,$years), but can be comparable to or even brighter than those of older USSNRs ($t\gtrsim 1000\,$years), especially for the `I\_SNR', `S\_SN', and `S\_SNR' models. In addition, we note that at later ages, the hot plasma can experience a compression from the expanding remnant, and the radio emission contribution from the plasma can be boosted further by this compression. Although the primary purpose of our study is on the modeling of USSNRs, the above discussion further advocates the importance of taking into account the CSM environment formed by the pre-SN mass loss activity of the progenitor in the USSNR emission model.}}

{
\subsection{Treatment of radiative cooling} \label{sec:radiativecooling}
Apart from the models presented so far, we have also performed extra simulations \deleted{only after the shock breakout }in which radiative cooling occurs in regions with a broader range of optical depths with $\tau < c/v$ to approximate the contribution of photon diffusion to the energy loss. While this approach overestimates the energy loss from radiative cooling, it is helpful nonetheless for assessing the robustness of our results. In these models with an enhanced energy loss, we found that the blastwave velocity is decreased by a few percent. This confirms that the impact of radiative cooling on the overall dynamics is small enough that it plays an insignificant role in the modeling of USSNRs.
}

\subsection{Effects of non-linear diffusive shock acceleration}~\label{sec:NLDSA}
{We have employed the} simplified treatment of particle acceleration and magnetic field amplification. In our study, non-linear effects in DSA are not considered, and the contribution of the pressure from cosmic-rays and its feedback to the hydrodynamics are not included. These effects can soften the energy distribution of accelerated electrons and could decrease the luminosity of non-thermal emission, including X-rays and gamma-rays \citep[e.g.,][]{2006ApJ...648L..33V,2019ApJ...876...27Y}. Our estimate of the USSNR population can thus be altered by including such effects (see Section~\ref{sec:population}). On the other hand, however, the dynamics of the USSNR blastwave is mainly determined by the distribution of the CSM. The lifetime of the blastwave is mainly limited by its interaction with the hot plasma in the vicinity of the ISM wall formed by the pre-SN mass loss. Thus, improving the treatment of the microphysics in shock acceleration plays a secondary role in the observable lifespan of a USSNR. 

\subsection{Parametrizations of $\epsilon_e$ and $E_{\rm min}$}~\label{sec:ee_Emin}
{There are two major simplifications in the parametrization for particle acceleration adopted in our study. 
First, some particle-in-cell simulations imply that the decrease of the Mach number (or the blastwave velocity) leads to a drop of the acceleration efficiency of protons \citep{2014ApJ...783...91C, 2018ApJ...864..105H}. This suggests the possibility that the acceleration efficiency of electrons also declines with a decreasing Mach number, while our study fixes $\epsilon_e$ at a constant value with time. Second, it is believed that electrons with momentum greater than $\sim \sqrt{m_e m_p}V_b$ follow a power-law distribution even below the relativistic regime. However, we have fixed the minimum energy of the power-law distribution at $E_{\rm min}$ in Equation (\ref{eq:NEnorm}) \citep[see also][]{2013ApJ...778..107S}. Hence, a decrease of the blastwave velocity leads to an increase of the number of electrons with a momentum $p_{\rm mom}$ within $\sqrt{m_e m_p}V_b \lesssim p_{\rm mom} \lesssim E_{\rm min}/c$. This effect is not included in our models. In summary, our study is over-estimating the radio luminosities and the actual brightness of the USSNRs could be fainter if the above two factors are accounted for. However, the blastwave velocity in our calculations is in the order of $\sim 10^9{\rm cm}\,$s\pow{-1} and the Mach number is sufficiently high in the young phase before the collision with the hot plasma. In the late phase ($t\gtrsim 1000\,$years), the blastwave dies away rapidly. Therefore, the system considered in our study is not prone to the situation described above. Moreover, even if we include the two effects mentioned above in our modeling, the resulted radio luminosities should be fainter than those reported in Section~\ref{sec:USSNRresult}, so that our conclusions on the characteristics and populations of USSNRs would not be affected qualitatively. {Furthermore, we have examined two values for $\epsilon_e$ shown in Table~\ref{tab:parameters}, and believe that the effect of the microphysics noted here can be investigated within this parameter space.}}

\subsection{Asphericity}~\label{sec:ashericity}
An aspherical configuration of the CE component and its effect on the wind hydrodynamics can be important as well. For instance it has been suggested that the material released by the CE ejection tends to distribute along the equatorial plane \citep{2019MNRAS.489.3334I}. Thus if the CE component resides in the vicinity of the SN progenitor it could affect the {subsequent} wind hydrodynamics. \deleted{However for USSNRs it is not the case for the reasons described below.} The gas ejected through the CE interaction should concentrate on the equatorial plane of the binary, while in the polar direction a static ISM should dominate. Then, the propagation of the wind driven by the RLO in the direction of the equatorial plane and the polar axis are regulated by the interaction {between the ISM with and without the CE component, respectively. Our simulations in Section~\ref{sec:CSMformation} show that the effect of the presence of the CE component is not significant regardless of the state of the ISM. From this point of view, by assuming a spherically blown wind from the progenitor binary, we can qualitatively speculate that the effect of possible non-spherical CE distributions would not be important.}

Besides, an anisotropy of the conformation of the wind can be expected to shape the non-spherical geometry of the CSM as proposed in the literature of Type IIn SNe \citep{2011AA...527L...6P, 2016ApJ...832..194K, 2019MNRAS.488.3089K}. It is worth investigating the multi-dimensional structures of the composed CSM taking into account the anisotropy of the circumstellar environment and the wind outflows. These aspherical configurations of the CSM can alter the properties of the radiation from the SNe or SNRs, which will be examined in detail in a future work \citep[see also e.g.,][]{2019AA...625A..24K, 2019ApJ...887..249S}.

\section{Summary}\label{sec:summary}

In this paper, we have investigated the characteristics of a SNR hosting a DNS binary, which we have termed a USSNR, using a grid of numerical models. A USSN has been proposed to be a transient event preceding the formation of a DNS binary. Before the USSN, the He star envelope is stripped away by the companion neutron star and escapes the binary system. By employing the mass-transfer history presented by \cite{2013ApJ...778L..23T}, we simulated the hydrodynamics of the wind expelled from the progenitor binary, and constructed the large-scale CSM structure around the USSN progenitor up to $\sim 100\,$pc. A hot plasma is formed in the vicinity of the ISM wall, which is found to play a critical role in governing the lifetime of the blastwave of the USSNR.

We also examined the dynamical and radiative evolution of a USSNR by considering a progenitor surrounded by the CSM composed by our simulation. We found that within the first $\sim 1000\,$years the blastwave traces the inner part of the CSM, producing a radio emission bright enough to be detected if the USSNR inhabits inside our Galaxy, though it is still fainter than those from typical SNRs. Once the blastwave collides with the hot plasma, it stalls rapidly and the radio luminosity also starts to decrease steadily. This dynamical behavior does not depend much on the strength of the CE ejection before the release of the helium gas from the progenitor binary. The surface brightness of the USSNR tends to be fainter than those of typical SNRs, while the diameter settles at $D\sim \mathcal{O}(10\,$pc$)$ similarly to the Galactic SNRs. Therefore, the USSNRs populate in the lower portion on the $\Sigma$-$D$ diagram compared to the observed Galactic SNRs, and this can serve as a useful diagnostics for the search of a USSNR. {We also confirmed that the initial ISM profile with a lower density allows the USSNR to expand further, leading to a lower surface brightness and a larger diameter.} Furthermore, we evaluated the observable lifespan of a USSNR to be $\sim 10^4\,$years, defined as the time interval from the explosion to the point when the radio luminosity has declined beyond the detection limit of the present radio surveys. Combining the short observable lifespan of the USSNRs with the small event rate of USSNe, we conclude that the expected number of active USSNRs is 
less than one out of the observed $10^{2-3}$ SNRs, which is consistent with the current non-detection of a SNR hosting a DNS.

\acknowledgments

The authors thank the anonymous referee for his or her fruitful suggestions, and Haruo Yasuda and Norita Kawanaka for their comments which have deepened the discussion in our study. T.M. acknowledges the support from the Iwadare Scholarship Foundation in the fiscal year 2020 and from Japanese Society for the Promotion of Science (JSPS) KAKENHI grant JP21J12145 since the fiscal year 2021. S.H.L. acknowledges support by JSPS grant No. JP19K03913 and the World Premier International Research Center Initiative (WPI), MEXT, Japan. K. M. acknowledges the support from JSPS KAKENHI grant 20H04737 and 18H05223. K. M. and T. J. M. acknowledge the support from JSPS KAKENHI grant 20H00174. T. T. acknowledges the support from JSPS KAKENHI grant JP17H06364, JP18H01212, and JP21H01088. T. J. M. acknowledges the support from JSPS KAKENHI grant JP18K13585, JP21K13966, and JP21H04997.

\appendix
\section{Derivation of $\rho_{\rm CE}$} \label{sec:app_rhoce}
In the simulation of the CSM formation, the value of $\rho_{\rm CE}$ must be specified to determine the initial density profile. We consider a CE component with a total mass $M_{\rm CE} = 10\Msun$ ejected into a static uniform ISM. The required condition is
\begin{eqnarray}
    \int_0^{R_\infty} 4\pi r^2(\rho(r)-\rho_{\rm ism})dr = M_{\rm CE},
\end{eqnarray}
where $R_\infty = 3\times 10^{21}$\,cm is the outermost radius of the simulation domain. {For the case $\rho(r) = \rho_{\rm CE} \exp(-r/R_{\rm CE}) + \rho_{\rm ism}$,} this can be analytically integrated, so that
\begin{eqnarray}
\rho_{\rm CE} \simeq \frac{M_{\rm CE}}{8\pi R_{\rm CE}^3} = 7.96\times 10^{-22}\,{\rm g cm}^{-3}
\end{eqnarray}
can be derived.

\section{Tests for the numerical code} \label{sec:numerics}
\begin{figure}[ht!]
\plotone{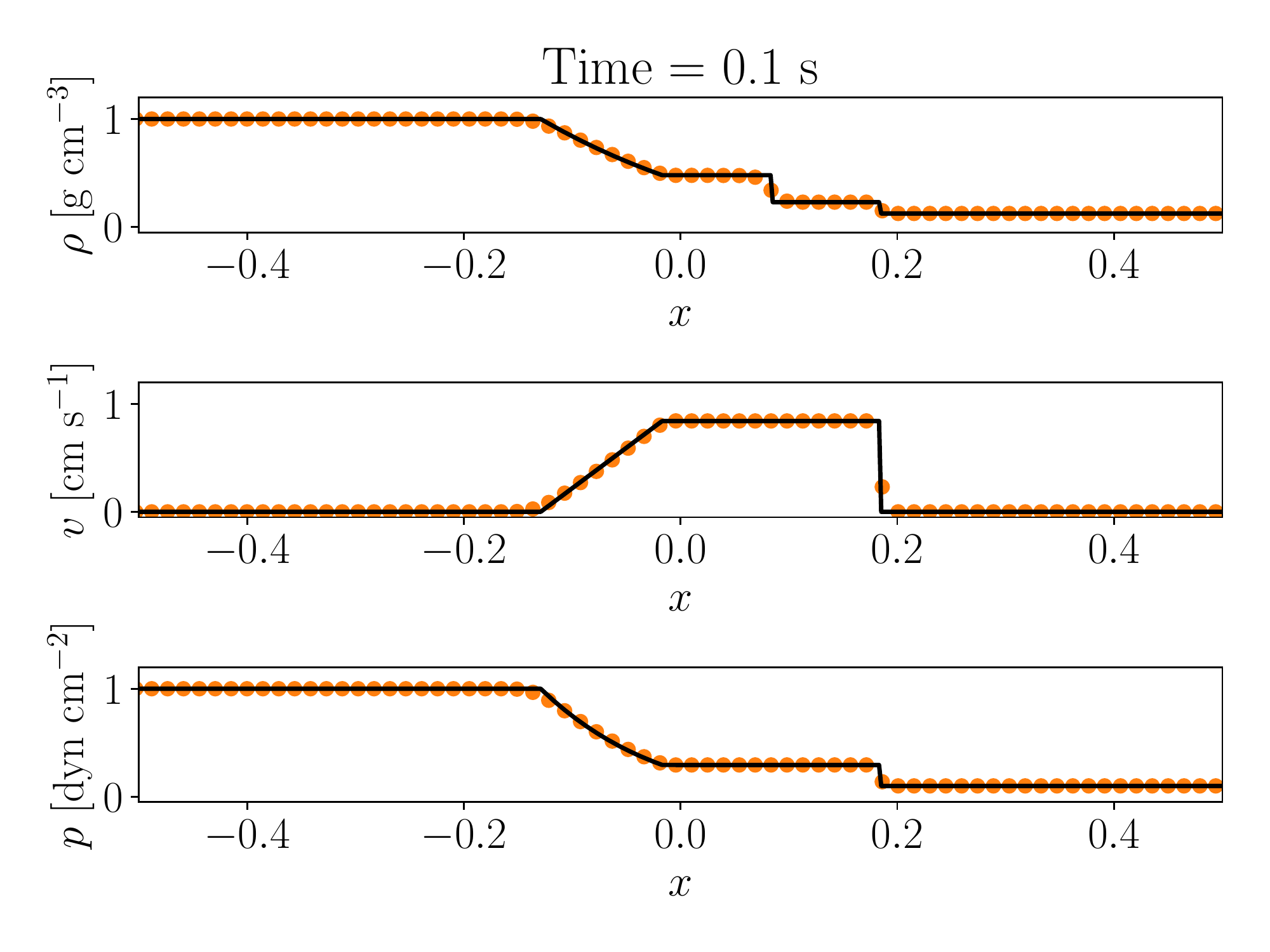}
\caption{The shock tube test. Black lines and orange circles show the exact solution and numerical solution derived by our code, respectively.}
\label{fig:app_shocktube}
\end{figure}

\begin{figure}[ht!]
\plotone{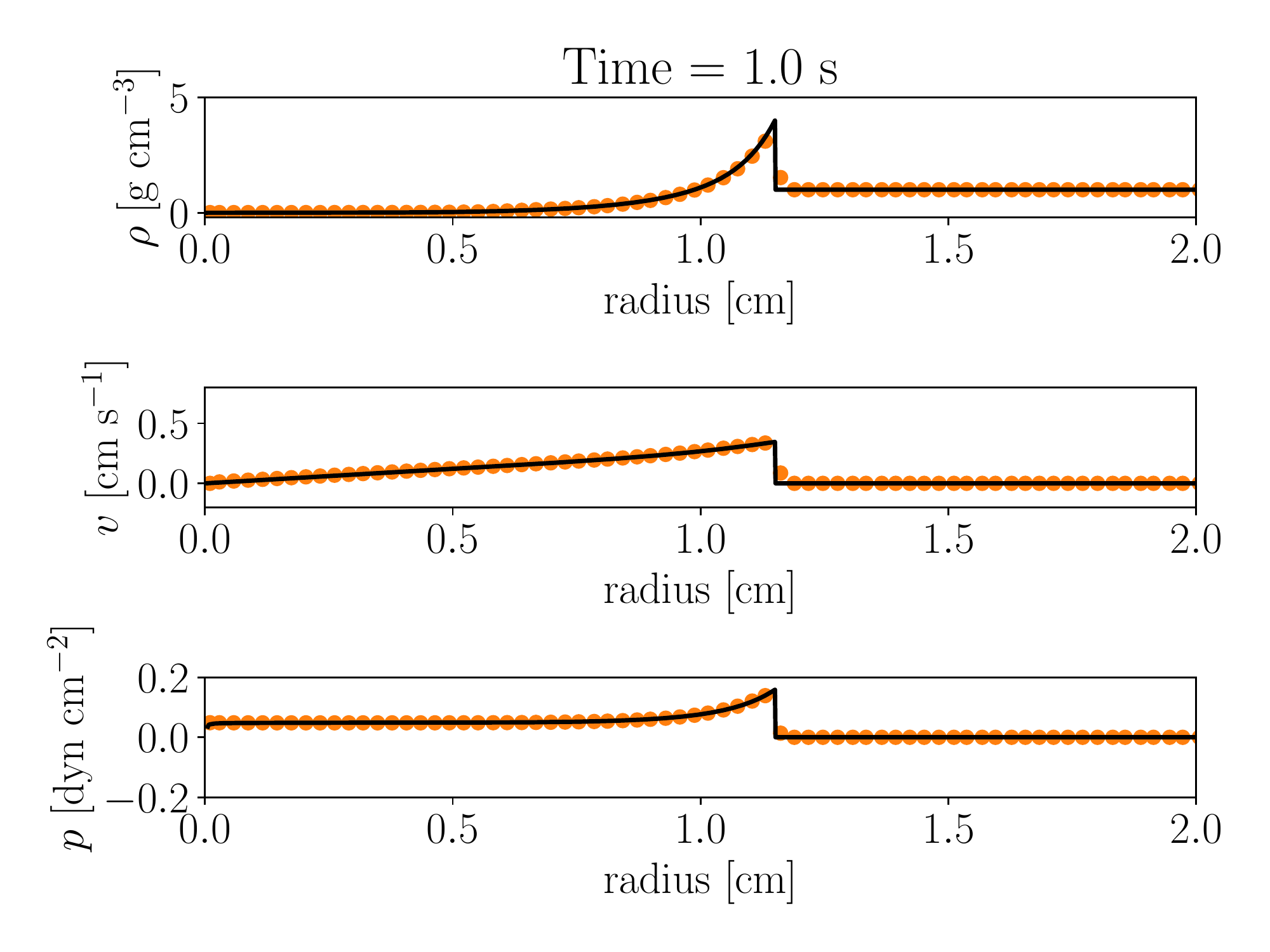}
\caption{The Sedov explosion test. The line and symbol have the same meaning as in Figure~\ref{fig:app_shocktube}.}
\label{fig:app_sedov}
\end{figure}

The numerical simulation code for the hydrodynamics employed in this study is verified in this section. Figure~\ref{fig:app_shocktube} shows the result of the shock tube problem with an adiabatic index $\gamma = 5/3$. At $t=0$, a static ($v=0$) gas is put into the simulation box, with a step function profile for its density and pressure centered at $x=0$ as follows; $\rho_L = 1.0, v_L = 0.0, p_L = 1.0, \rho_R =0.125, v_R =0.0, p_R = 0.1$ (the subscripts $L$ and $R$ denote $x<0$ and $x\geq 0$, respectively). The numerical solution successfully reproduces the profiles given by the exact solution. Furthermore, Figure~\ref{fig:app_sedov} displays the Sedov solution at $t=1.0$ second in which an explosion energy $E_{\rm sedov} = 1$ erg is deposited into a uniform medium with $\rho_{\rm sedov} = 1.0\times 10^{-24}$ g cm\pow{-3} \citep[][]{1959sdmm.book.....S}. The results are again in a good agreement with the analytical solutions for the density, velocity, and pressure profiles, as well as a good match of the shock radius given by $R=1.15(E_{\rm sedov}t^2/\rho_{\rm sedov})^{0.2}$. These two experiments assure us a good accuracy of our numerical code.

\newpage

\newpage

\bibliography{manuscript}{}
\bibliographystyle{aasjournal}

\begin{turnpage}
\begin{deluxetable}{cccccc}[ht!]
\label{tab:LCplots}
\tablecaption{Samples of the observed Galactic core-collapse SNRs}
\tablehead{
\colhead{SNR name} &
\colhead{Common name} &
\colhead{Age [yr]} &
\colhead{Distance [kpc]} &
\colhead{Flux density [Jy]} &
\colhead{References}
}
\startdata
G15.9+0.2 & &1000 - 3000 & 8.5 - 16.7 & 5.0 &  \cite{2018MNRAS.479.3033S} \\
G34.7-0.4 & W44 & 7900 - 8900 & 2.7 - 3.3 & 240 & \cite{2013Sci...339..807A, 2012PASJ...64..141U} \\
G43.3-0.2 & W49B & 2900 - 6000 & 10.9 - 11.7 & 38 & \cite{1994ApJ...437..705M, 2014ApJ...793...95Z}\\
G67.7+1.8 & & 5000 - 13000 & 7.0 - 17.0 & 1 & \cite{2009AA...494.1005H}\\
G111.7-2.1 & CasA & 316 - 352 & 3.3 - 3.7 & 2400 & \cite{2003ApJ...589..818D}\\
G189.1+3.0 & IC 433 & 3000 - 30000 & 0.7 - 2.0 & 165 & \cite{2013Sci...339..807A, 2017MNRAS.472...51A} \\
G260.4-3.4 & Puppis A & 3700 - 4500 & 1.3 - 2.2 & 130 & \cite{2017MNRAS.464.3029R} \\
G266.2-1.2 & Vela Jr & 2400 - 5100 & 0.5 - 1.0 & 50 & \cite{2015ApJ...798...82A} \\
G291.0-0.1 & & 1300 - 10000 & 3.5 - 6.0 & 16 & \cite{1986MNRAS.219..815R}\\
G292.0+1.8 & & 2930 - 3050 & 5.3 - 7.1 & 15 & \cite{2003ApJ...594..326G}\\
G296.1-0.5 & & 2800 - 28000 & 2.0 - 4.0 & 8 & \cite{1996AAS..118..329W, 2012MNRAS.419.1603G} \\
G308.4-1.4 & & 5000 - 7500 & 9.1 - 10.7 & 0.4 &  \cite{2012AA...544A...7P} \\
G309.2-0.6 & & 700 - 4000 & 2.0 - 6.0 & 7 & \cite{1998MNRAS.299..812G, 2001ApJ...548..258R} \\
G330.2+1.0 & & 1000 - 3000 & 4.6 - 5.2 & 5 & \cite{1996AAS..118..329W}\\
G347.3-0.5 & & 1624 - 1626 & 0.5 - 1.6 & 30 & \cite{2001ApJ...563..191E, 2003PASJ...55L..61F}\\
G350.1-0.3 & & 600 - 1200 & 4.5 - 9.0 & 6 & \cite{2011ApJ...731...70L} \\
\enddata
\tablecomments{The values of the radio flux are cited from \cite{2017yCat.7278....0G}. Notice that the explosion types of some of these samples have not been clarified. For details, see also Table 3 in \cite{2019ApJ...876...27Y}.}
\end{deluxetable}
\end{turnpage}

\end{document}